\documentclass[10pt]{article}      
\usepackage{fullpage}

\usepackage{amsmath,amssymb,amsfonts,amsthm,amscd} 
\usepackage{graphics}                 

\usepackage{tensor}                   

\usepackage{color}                    
\usepackage{mdframed}                 
\usepackage{mathrsfs}
\usepackage{dsfont}
\usepackage{bbold}
\usepackage{url}         
\usepackage{mathtools} 
\usepackage[toc,page]{appendix} 
\usepackage{empheq}
\usepackage{cancel}                     
\usepackage{longtable} 
\usepackage{soul} 

\usepackage{enumitem}    

\usepackage{graphicx,subcaption}
\usepackage{wrapfig}
\usepackage[subfigure]{tocloft}

\usepackage{nccmath}   
\usepackage{hyperref}
\hypersetup{
  colorlinks   = true, 
  urlcolor     = black, 
  linkcolor    = blue, 
  citecolor   = blue 
}

\newcommand\Item[1][]{%
  \ifx\relax#1\relax  \item \else \item[#1] \fi
  \abovedisplayskip=0pt\abovedisplayshortskip=0pt~\vspace*{-\baselineskip}}
\usepackage[textsize=tiny]{todonotes}

\usepackage{tikz}
\usetikzlibrary{arrows,decorations.markings,patterns}
\usetikzlibrary{automata}
\usetikzlibrary{positioning}  
\usetikzlibrary{arrows}       
\tikzset{node distance=3.5cm, 
         every state/.style={ 
           thick,
           fill=gray!10},
         initial text={},     
         double distance=4pt, 
         every edge/.style={  
         draw,
           auto,
           semithick}}

\theoremstyle{plain}

\theoremstyle{remark}
\AtEndEnvironment{remark}{\qed}

\numberwithin{equation}{section}

\definecolor{aliceblue}{rgb}{0.94, 0.97, 1.0} 
\definecolor{azuremist}{rgb}{0.94, 1.0, 1.0} 
\definecolor{antiquewhite}{rgb}{0.98, 0.92, 0.84} 
\definecolor{ivory}{rgb}{1.0, 1.0, 0.94}

\newcommand\bra[1]{\left({#1}\right)}

\newcommand{\In}{\mathrm{in}}
\newcommand{\Out}{\mathrm{out}}

\newcommand{\sym}{\mathrm{sym}}
\newcommand{\asym}{\mathrm{asym}}

\newcommand\pra[1]{\left[{#1}\right]}

\DeclareMathOperator{\Grad}{grad}

\DeclareMathOperator{\JJ}{\mathbb{J}}

\DeclareMathOperator{\Prob}{\mathrm{Prob}}
\DeclareMathOperator*{\sumsum}{\sum\!\sum}
\newcommand{\sumxly}{\sumsum_{\substack{x,y\in\X\\ x<y}}}
\newcommand{\sumxny}{\sumsum_{\substack{x,y\in\X\\ x\neq y}}}
\newcommand{\super}[1]{^{\scriptscriptstyle{(#1)}}}

\def\tp{^\mathsf{T}}
\DeclareMathOperator{\diag}{ \mathop{\mathrm{diag}} }

\newcommand\ghost{\odot}

\def\E{\mathcal{E}}
\def\H{\mathcal{H}}
\def\L{\mathcal{L}}

\def\P{\mathcal{P}}

\def\U{\mathcal{U}}
\def\V{\mathcal{V}}

\def\X{\mathcal{X}}

\def\Q{\mathcal{Q}}
\def\RR{\mathbb{R}}
\def\R{\mathbb{R}}

\def\NN{\mathbb{N}}

\let\div\relax

\DeclareMathOperator{\grad}{\nabla}
\DeclareMathOperator{\div}{div} 

\DeclareMathOperator{\dgrad}{ \overline{\scalebox{0.9}{\ensuremath\nabla}} }
\DeclareMathOperator{\ddiv}{\overline{\scalebox{0.9}{\textup{div}}}}

\newcommand{\Ain}{{\In A}}
\newcommand{\Aout}{{\Out A}}
\newcommand{\Bin}{{\In B}}
\newcommand{\Bout}{{\Out B}}
\newcommand{\Cin}{{\In C}}
\newcommand{\Cout}{{\Out C}}

\newcommand{\Edge}{\mathcal{E}} 

\makeatletter
  \newcommand{\eqnum}{\leavevmode\hfill\refstepcounter{equation}\textup{\tagform@{\theequation}}} 
\makeatother

\allowdisplaybreaks

\date \today
\title{
 Untangling Dissipative and Hamiltonian effects in bulk and boundary driven systems 
}
\author{D.R. Michiel Renger\thanks{Department of Mathematics, Technische Universit\"at M\"unchen, Boltzmannstr.\ 3, 85748 Garching, Germany. 
Email: \href{mailto:d.r.m.renger@tum.de}{d.r.m.renger@tum.de}
} \
and Upanshu Sharma\thanks{School of Mathematics and Statistics, The University of New South Wales, Sydney 2052, Australia.\ Email: \href{mailto:upanshu.sharma@unsw.edu.au}{upanshu.sharma@unsw.edu.au}}}

\begin{document}
\mathtoolsset{centercolon} 
\maketitle

\begin{abstract}
Using the theory of large deviations, macroscopic fluctuation theory provides a framework to understand the behaviour of non-equilibrium dynamics and steady states in \emph{diffusive} systems. We extend this framework to a minimal model of non-equilibrium \emph{non-diffusive} system, specifically an open linear network on a finite graph. We explicitly calculate the dissipative bulk and boundary forces that drive the system towards the steady state, and non-dissipative bulk and boundary forces that drives the system in orbits around the steady state. Using the fact that these forces are orthogonal in a certain sense, we provide a decomposition of the large-deviation cost into dissipative and non-dissipative terms. We establish that the purely non-dissipative force turns the dynamics into a Hamiltonian system. These theoretical findings are illustrated by numerical examples. 
\end{abstract}

\section{Introduction}

It is well known that if a microscopic stochastic particle system is in detailed balance, then large fluctuations around the macroscopic dynamics (large-deviations theory) induce a gradient flow of the free energy. This principle was first discovered by Onsager and Machlup in their groundbreaking paper~\cite{Onsager1953I} for a simple process with vanishing white noise, and their result may be identified with the more rigorous and general Freidlin-Wentzell theory~\cite{FreidlinWentzell2012}. However, as Onsager and Machlup stated in 1953:
\begin{quote}
The proof of the reciprocal relations \lbrack...\rbrack\,was based on the hypothesis of microscopic reversibility, which we retain here. This excludes rotating systems (Coriolis forces) and systems with external magnetic fields. The assumption of Gaussian random variables is also restrictive: Our system must consist of many ``sufficiently'' independent particles, and equilibrium must be stable at least for times of the order of laboratory measuring times.
\end{quote}
Regarding the Gaussian noise, extensions to different noise have been known for a long time, see for example~\cite{Bertini2004minimum}. What these models have in common is that, although on a microscopic level the (vanishing) noise is non-Gaussian, macroscopically these systems are diffusive. Therefore, the large deviations corresponding to this hydrodynamic limit have a quadratic rate functional (often called the `dynamical action'), i.e.\ of the form $\frac14\int_0^T\!\lVert\dot\rho(t)+\Grad\V(\rho(t))\rVert^2_{\rho(t)}\,dt$ for some $\rho$-dependent norm on velocities, gradient corresponding to that norm, and free energy or quasipotential $\V$. Here $\rho(t)$ is usually the (hydrodynamic) particle density of an underlying microscopic stochastic particle system, for instance as in lattice gas models~\cite{BDSGJLL2002,Bertini2003SSEPbdd} or interacting SDEs~\cite{DawsontGartner87}, but $\rho$ could also be the local energy or temperature as in the Kipnis-Marchioro-Presutti model of heat conduction~\cite{BertiniGabrielliLebowitz2005KMP}.
Expanding the squares in the quadratic rate functional and applying a chain rule then yields the form
$\frac14\int_0^T\!\lVert\dot\rho(t)\rVert^2_{\rho(t)}\,dt + \int_0^T\!\lVert\frac12\delta\V(\rho(t))/\delta\rho\rVert^2_{\rho(t)*}\,dt+\frac12\V(\rho(T))-\frac12\V(\rho(0))$, as predicted by Onsager and Machlup, where $\|\cdot\|_{\rho*}$ denotes the dual norm on forces or potentials. As the number of particles increases, the noise vanishes, the rate functional becomes $0$ and these terms represent a free energy balance, corresponding to the dissipative system or gradient flow $\dot\rho(t)=-\Grad\V(\rho(t))$. Different, for example, Poissonian noise may lead to non-quadratic large deviations, but as discovered in~\cite{MielkePeletierRenger14}, the Onsager-Machlup principle still holds for systems in detailed balanced if one allows for \emph{nonlinear} macroscopic response relation between the force $-\frac12\delta\V(\rho)/\delta\rho$ and the velocity $\dot\rho$.

Regarding systems with additional (non-dissipative) `rotating' effects mentioned by Onsager and Machlup, these correspond to thermodynamically open systems, which can for example be physically realised by coupling with separate heat baths, or by injecting and extracting matter at boundaries, while microscopically they correspond to a breaking of detailed balance. These \emph{non-equilibrium systems} are particularly challenging due to the combination of dissipative and non-dissipative effects that are strongly intertwined.

The field of macroscopic fluctuation theory (MFT)~\cite{BDSGJLL2002,Bertini2004minimum,BDSGJLL2015MFT} allows an orthogonal decomposition into dissipative and non-dissipative dynamics, albeit, for diffusive systems. The non-dissipative part of the dynamics is represented by forces that cause rotating, possibly divergence-free motion, so that a free energy balance as above is bound to fail unless one takes particle fluxes $j$ into account. For diffusive systems, this yields a large-deviation rate functional of the form
$
  \frac14\int_0^T\!\lVert j(t)+\Grad\V(\rho(t))+\chi(\rho(t)) A \rVert^2_{\rho(t)}\,dt
$,
$\dot\rho(t)=-\div j(t)$
for some norm on fluxes and corresponding gradient, free energy $\V$ and divergence-free vector field $A$, and mobility $\chi(\rho)$ that transforms forces into fluxes~\cite{BDSGJLL2015MFT}. In MFT one then exploits the fact that the dissipative force $-\frac12\grad\delta\V(\rho)/\delta\rho$ and non-dissipative force $-\frac12A$ are orthogonal in the dual norm $\|\cdot\|_{\rho*}$ on forces, which allows the rate functional to be written as \cite{BDSGJLL2015MFT}
\begin{multline}\label{eq:MFT decomp}
  \frac14\int_0^T\!\lVert j(t)\rVert^2_{\rho(t)}\,dt 
  +
  \int_0^T\!\lVert \tfrac12\grad\delta\V(\rho(t))/\delta\rho\rVert^2_{\rho(t)*}\,dt
  +
  \frac12\V(\rho(T))-\frac12\V(\rho(0))
  +
  \int_0^T\!\lVert\tfrac12 A \rVert^2_{\rho(t)*}\,dt
  +
  \int_0^T\!\langle \tfrac12A,j(t)\rangle\,dt.
\end{multline}

To highlight the importance of such a decomposition, we discuss three different interpretations of the full large-deviation rate functional. In the first, mathematical interpretation, the large-deviation rate functional characterises the exponential probability decay in the zero-noise limit, for untypical paths $(\rho(t),j(t))$ that deviate from the macroscopic dynamics (see Section~\ref{sec:model}). The first four terms in~\eqref{eq:MFT decomp} are the flux version of the Onsager-Machlup decomposition, and therefore represent the convergence speed of the dissipative part of the system, that is, when the system \emph{would} have been in detailed balance. The last two terms are an additional decay rate that is purely caused by the rotating force.

In the second interpretation, the large-deviation rate functional corresponds to the free energy that must be injected in the system in order to create a path $(\rho(t),j(t))$ that deviates from the typical macroscopic dynamics. The last two terms are the Fisher information/dissipation and work done by the rotating force; together they represent an additional contribution to the free energy cost.

In the last interpretation we look at the typical behaviour. A maximised probability corresponds to a minimised rate functional, so setting the rate functional to $0$, we obtain a non-equilibrium free energy balance as above. Instead of looking at convergence speed as the noise vanishes, we now look at convergence of the macroscopic system to its steady state as $T\to\infty$. This can be characterised by the free energy loss $\frac12\V(\rho(T))-\frac12\V(\rho(0))$, for which we now obtain an explicit expression, with explicit and distinct contributions due to the dissipative and non-dissipative (rotating) forces in the system.

If in addition, the non-dissipative force has a Hamiltonian structure, then connections to GENERIC can be made~\cite{Ottinger2005,KLMP2020math,PRS2021TR}. However, this connection is a feature of the fact that the large deviations are quadratic~\cite{RengerSharmaGENERICTR}, which fails for the systems that we study in this paper.

For more general systems, the flux large deviations are of the form $\int_0^T\!\L(\rho(t),j(t))\,dt$, but for non-diffusive systems this action $\L(\rho,j)$ is not quadratic in the flux $j$. Although lacking a natural notion of orthogonality, it is recently discovered that the dissipative force $-\tfrac12\grad\delta V(\rho(t))/\delta\rho$ and non-dissipative force $F^\asym$ are orthogonal in some generalised sense, allowing a decomposition similar to \eqref{eq:MFT decomp} \cite{KaiserJackZimmer2018,RengerZimmer2021, PRS2021TR}
\begin{multline}
    \int_0^T\!\L(\rho(t),j(t))\,dt  
    =\int_0^T\!\Psi(\rho(t),j(t))\,dt+\int_0^T\!\Psi^*_\circledast\big(\rho(t),-\tfrac12\grad\delta V(\rho(t))/\delta\rho\big)\,dt+\frac12\V(\rho(T))-\frac12\V(\rho(0)) \\ + \int_0^T\!\Psi^*_\circledast(\rho(t),F^\asym)\,dt - \int_0^T\!F^\asym\cdot j(t)\,dt.
\label{intro:decomp}
\end{multline}
Here the \emph{dissipation potential} $\Psi$ generalises the squared norm on fluxes, $\Psi^*$ is its convex dual, and one of the two dual potentials $\Psi^*_\circledast$ needs to be modified, see Sections~\ref{sec:forces} and \ref{sec:decom} for details.

One significant implication of \eqref{intro:decomp} is an \emph{explicit} expression~\eqref{eq:antisym work} for the non-dissipative work $\int_0^T\! F^\asym\cdot j(t)\,dt$ along the macroscopic zero-cost dynamics $\L(\rho,j)=0$. Similarly we shall derive an explicit expression~\eqref{eq:free energy loss} for the dissipative work or free energy loss $\frac12\V(\rho(T))-\frac12\V(\rho(0))$. Both expressions have separate contributions due to dissipative and non-dissipative forces, and both are non-positive. Together with the non-positivity of the total work $\int_0^T\!F(\rho)\cdot j(t)\,dt$, this coincides with what is sometimes called the `three faces of the second law'; for chemical reaction networks these three signs have been derived in \cite{FreitasEsposito2022}. Similar bounds can also be found in~\cite{Maes2017a,Maes2018}. However we point out that the above decomposition holds for any path, not just along the zero-cost dynamics, and we extend the analysis to boundary-driven systems. 

Decompositions of the type~\eqref{intro:decomp} have been studied in depth in \cite{PRS2021TR}, but the details are only known for a few models, in particular models that are driven out of equilibrium by bulk effects. 

In the current work our aim is to precisely understand how bulk and boundary effects can jointly drive a system out of detailed balance, and we achieve this by studying a linear network with open boundaries. This minimal model is sufficiently rich to understand the role of bulk and boundary individually and provide guidelines to more complex nonlinear systems.

Our novel contributions are twofold. First, we calculate the flux large deviations for the linear network with open boundaries and explicitly calculate all boundary-bulk dissipative--non-dissipative forces and corresponding dissipation potentials and Fisher informations, from we derive the decomposition~\eqref{intro:decomp}. In order to do so we explicitly calculate the free energy/quasipotential $\V$ as the large-deviation rate for the invariant measure.

The second contribution lies in the study of purely non-dissipative systems ($\V=0$) as a counterpart to dissipative gradient flows ($F^\asym=0$). For a few bulk-driven models, it was recently discovered that 
such dynamics in fact correspond to a Hamiltonian system with periodic orbit solutions \cite{PRS2021TR}. This precisely delineates the role of dissipative forces which drive the system to its steady state from non-dissipative forces which drive the system out of detailed balance precisely through a Hamiltonian flow.
In the current paper we show that for our boundary-bulk driven model, the purely non-dissipative dynamics are indeed a Hamiltonian system, and we explicitly calculate the conserving energy and Poisson structure, and show that the Poisson structure indeed satisfies the Jacobi identity.

\paragraph{Terminology.} We avoid the words (non)equilibrium and (ir)reversibility and talk about (non)detailed balance instead. The distinction between boundary and bulk refers to large graphs with many internal nodes between which particles can hop, and only a few boundary nodes where particles can also be injected or extracted from the system. However, it turns out to be notationally convenient to assume that injection and extraction may in principle occur at any node in the system.

\section{Model}
\label{sec:model}

\begin{figure}[htb]
\centering
\begin{tikzpicture}[>=stealth]
\tikzstyle{every node}=[font=\scriptsize];

\node[state, fill=antiquewhite] (B) {B};
\node[state, below right of=B, fill=antiquewhite] (C) {C};
\node[state, below left of=B, fill=antiquewhite] (A) {A};


\draw (B) edge[->,bend left=12] node[above,pos=0.5,sloped] {$Q_{BC}$}  (C);
\draw (A) edge[->,bend left=12] node[above,pos=0.5,sloped] {$Q_{AB}$} (B);
\draw (C) edge[->,bend left=12] node[below,pos=0.5,sloped] {$Q_{CB}$} (B);
\draw (A) edge[->,bend left=12] node[above,pos=0.5,sloped] {$Q_{AC}$} (C);
\draw (B) edge[->,bend left=12] node[below,pos=0.4,sloped] {$Q_{BA}$} (A);
\draw (C) edge[->,bend left=12] node[below,pos=0.5,sloped] {$Q_{CA}$} (A);

\draw[semithick,->] (C) to  [bend left=10] node[above] {$\lambda_{\Cout}$} (4.5,-3) ;
\draw[semithick,<-] (C) to [bend right=10] node[below] {$\lambda_{\Cin}$}(4.3,-3.2);

\draw[semithick,dotted,->] (B) to [bend left=10] node[left,midway] {$\lambda_{\Bout}=0$}   (-0.1,1.2);
\draw[semithick,dotted,<-] (B) to  [bend right=10] node[right,midway] {$\lambda_{\Bin}=0$} (0.1,1.2);

\draw[semithick,<-] (A) to [bend right=10] node[above] {$\lambda_{\Ain}$} (-4.5,-3);
\draw[semithick,->] (A) to [bend left=10] node[below] {$\lambda_{\Aout}$} (-4.3,-3.2);


\end{tikzpicture}
\caption{An example of a linear network with open boundaries. }
\label{fig:open-bound-chain}
\end{figure}
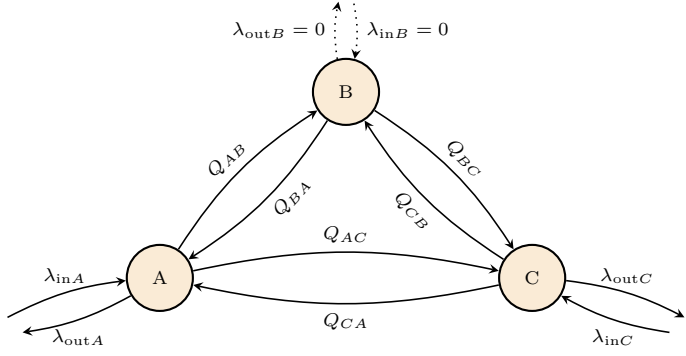

We consider a large number of particles hopping between different nodes on a finite graph $\X$, where particles may also be removed or injected. The nodes may be interpreted as spatial compartments, or more abstractly as chemical species/states in the case of unimolecular reactions or discrete protein folding. The rate at which a particle hops between nodes $x$ and $y$ is denoted as $Q_{xy}$, and $\lambda_{\In x},\lambda_{\Out x}$ are the rates at which particles are added and removed respectively from node $x$ on the graph.
See Figure~\ref{fig:open-bound-chain} for an example and Section~\ref{sec:numerics} for numerical results for this example. We only assume (a) $Q_{xy}=0\iff Q_{yx}=0$, (b) $\lambda_{\In x}=0\iff\lambda_{\Out x}=0$ and (c) that the graph with nonzero weights $Q_{xy}>0$ is irreducible.

Defining $Q_{xx}:=-\sum_{y\in\X,y\neq x} Q_{xy}$ as usual, the macroscopic evolution of mass $\rho(t)\in\RR^\X$  on the graph is 
\begin{equation}\label{eq:ODE}
    \dot\rho(t)=\big(Q - \diag(\lambda_\Out)\big)\tp \rho(t) + \lambda_\In.
\end{equation}

This model is different from the usual bulk-boundary systems studied in the literature (see~\cite[Section VIII]{BDSGJLL2015MFT} for a comprehensive list). First, we make the choice of dealing with independent particles to simplify the ensuing analysis on large-deviations and decompositions of the large-deviation cost. While we do expect similar ideas to hold for interacting particles (such as stochastic chemical-reaction networks with boundaries) this is left to future work. Second, we make the atypical choice of allowing particle creation/annihilation at each node of the graph. The classical setting where a large bulk has a few boundary nodes is a special case of our model. This is seen for instance by adding many intermediate nodes in Figure~\ref{fig:open-bound-chain} with zero in-out flow, i.e.\   $\lambda_{\mathrm{in}}=\lambda_{\mathrm{out}}=0$ but non-zero edge weights. Note that the discussions and calculations in the rest of the paper do not change if one chooses to work with fewer boundary nodes as opposed to the current setup wherein every node is a boundary node. We make this choice for simplicity of notation since choosing a few boundary nodes would require us to distinguish between bulk/boundary nodes in every ensuing summation.

In order to investigate non-dissipative effects we study \emph{net} fluxes $j(t)$ in addition to the mass density $\rho(t)$. To this aim we equip the graph $\X$ with an (arbitrary) ordering, which defines the positive edges $\Edge:=\{(x,y): x,y \in\X, x<y\}\cup\{(\In x): x\in \X\}$.
The macroscopic \emph{flux formulation} of~\eqref{eq:ODE} is
\begin{align}\label{eq:hydroLim}
  &j_{xy} (t) = j^0_{xy}(\rho(t)),                &&
  &j_{\In x}(t)=j^0_{\In x}(\rho(t)),             &&
  &\dot \rho_x(t)=-\ddiv_x j(t),                   &&
\end{align}
where the zero-cost flux on the positive edges is
$
  j^0_{xy}(\rho):= \rho_xQ_{xy}-\rho_y Q_{yx}
$
and
$
  j^0_{\In x}(\rho):=\lambda_{\In x} - \rho_x\lambda_{\Out x}
$.
The discrete divergence operator $\ddiv:\RR^\E\to\RR^\X$ on fluxes  is defined as
\begin{equation}\label{eq:disDiv}
  \ddiv_x j := \sum_{y\in\X: y>x} j_{xy} -\sum_{y\in\X:y<x} j_{yx} -j_{\In x}.
\end{equation}
The first two sums define the classic discrete divergence for closed systems; together with the last term, $\ddiv_x j$ describes the net flow out of a node $x$ for open systems. In particular $\sum_x \ddiv_x j=-\sum_x j_{\In x}$ equals the total net flow out of the system. 
This particular definition of the discrete divergence accounts for the net fluxes and arises from the following natural underlying (stochastic) microscopic particle system.

The large parameter $n$ will be used to control the \emph{order} of the total number of particles in the system, although this number is generally not conserved over time.
At each node $x$, new particles are randomly created with rate $n\lambda_{\In x}$ and independently of all other particles, each particle either randomly jumps to node $y$ with rate $Q_{xy}$, or is randomly destroyed with rate $\lambda_{\Out x}$. We are interested in the random particle density $n\rho\super{n}_x(t)$ which counts the number of particles at node $x$ and time $t$, the cumulative net flux $n W\super{n}_{xy}(t)$, which counts the number of jumps $x\to y$ minus the jumps $y\to x$ in time interval $(0,t\rbrack$, and the net flux $n W\super{n}_{\In x}$, counting the number of particles created minus the number of particles destroyed at that node $x$ in time interval $(0,t\rbrack$.

By Kurtz' Theorem~\cite{Kurtz1970a}, the Markov process $(\rho^{(n)}(t),W^{(n)}(t))$ converges 
as $n\to\infty$ to the solution $(\rho(t),w(t))$ of~\eqref{eq:hydroLim}, where we identify the derivative $\dot w(t)$ of the cumulative net flux with the net flux $j(t)$. We stress that for finite $n$, the continuity equation $\dot\rho=-\ddiv j$ holds almost surely, but random fluctuations occur in the fluxes.

On an exponential scale, these fluctuations satisfy a large-deviation principle~\cite{ShwartzWeiss1995,MaesNetocnyWynants2007Markov,Renger2018a,PattersonRenger2019}\footnote{We ignore possible contributions from random initial fluctuations since they they play no role in our paper whatsoever.}
\begin{equation}\label{eq:LDP}
  \Prob\bigl( (\rho\super{n},W\super{n})\approx (\rho,w)\bigr) \stackrel{n\to\infty}{\sim} \exp\bigl(\textstyle{-n\int_0^T\!\L\bigl(\rho(t),\dot w(t)\bigr)\,dt}\bigr),
\end{equation}
where we implicitly set the exponent to $-\infty$ if the continuity equation $\dot\rho(t)\equiv-\ddiv j(t)$ is violated, and\footnote{In the case when only a few nodes are boundary nodes, i.e.\ only a subset of all nodes satisfy $\lambda_{\In},\lambda_{\Out}\neq 0$, the second summation in~\eqref{def:Lag} is only taken over those boundary nodes.}
\begin{equation}\label{def:Lag}
\begin{aligned}
  \L(\rho,j) &:= 
    \sumxly \inf_{j^+_{xy}\geq0} \bigl[ s(j^+_{xy} \mid \rho_x Q_{xy}) + s(j^+_{xy}-j_{xy} \mid \rho_y Q_{yx})\bigr]  \\
    &\qquad\quad + \sum_{x\in\X} \inf_{j^+_{\In x}\geq0} \bigl[   s(j^+_{\In x} \mid \lambda_{\In x}) + s(j^+_{\In x}-j_{\In x} \mid \rho_x \lambda_{\Out x})\bigr],
\end{aligned}
\end{equation}
using the usual (non-negative and convex) relative entropy function $s(a\mid b):=a\log\frac{a}{b} - a + b$. 
The infima in the definition of $\L$ contracts the large-deviation principle of the one-way fluxes to the large-deviation principle of the net fluxes~\cite[Thm.~4.2.1]{DemboZeitouni09}. 
Note that $\L$ is non-negative  and satisfies $\L(\rho,j^0)=0$, i.e.\ $j^0$ is the zero-cost flux, since $s(a\mid b)=0$ if and only if $a=b$. The optimal one-way fluxes in~\eqref{def:Lag} are given by
\begin{align}
    j^+_{xy} = \mfrac12 j_{xy}+\sqrt{\mfrac14 j_{xy}^2+\rho_xQ_{xy}\rho_yQ_{yx}},
    &&\text{and}&&
    j^+_{\In x} =  \mfrac12 j_{\In x}+\sqrt{\mfrac14 j_{\In x}^2+\lambda_{\In x} \rho_x\lambda_{\Out x}},
\end{align}
which yields an explicit but less insightful expression for the cost function~\eqref{def:Lag}.

It will often be convenient to work with the convex (bi-)dual of $\L(\rho,\cdot)$, defined for forces $\zeta\in\R^\Edge$ acting on net fluxes
\begin{align}\label{def:Ham}
&\H(\rho,\zeta):=\sup_{j\in\RR^{\Edge}} \zeta\cdot j - \L(\rho,j)\\
              &= \sumxly \pra{\rho_x Q_{xy}\bra{e^{\zeta_{xy}}-1}+\rho_y Q_{yx}\bra{e^{-\zeta_{xy}}-1} }+ \sum_{x\in\X}\pra{ \lambda_{\In x} \bra{e^{\zeta_{\In x}}-1}+ \rho_x \lambda_{\Out x}\bra{e^{-\zeta_{\In x}}-1} }.\notag
\end{align}
As $j^0(\rho)$ from~\eqref{eq:hydroLim} is the zero-cost flux, we can write $j^0(\rho)=\grad_\zeta\H(\rho,0)$.

\section{Invariant measure, quasipotential and time reversal}

The macroscopic equation~\eqref{eq:ODE} has a unique, coordinate-wise positive steady state $\pi\in\R^\X$ (see Appendix~\ref{app:steady-state}). Moreover, for fixed $n$, the random process $\rho\super{n}(t)$ has the unique invariant measure $\Pi\super{n}\in \P(\RR^\X)$ of product-Poisson form 
\begin{equation}\label{eq:product-Poisson}
  \Pi\super{n}(\rho):=
  \begin{cases}
      \prod_{x\in\X} \frac{(n\pi_x)^{n\rho_x}e^{-n\pi_x}}{(n\rho_x)!}
      &\rho\in(\tfrac1n\NN_0)^\X,\\
    0,
      &\text{otherwise},
  \end{cases}
\end{equation}
see Appendix~\ref{app:steady-state}.
By Stirling's formula one obtains that the invariant measure satisfies a large deviation principle $\Pi\super{n}(\rho)\sim e^{-n\V(\rho)}$ with quasipotential
\begin{equation}\label{eq:QP}    
    \V(\rho):= \sum_{x\in\X} s(\rho_x|\pi_x),
\end{equation}
which can also be interpreted as $(k_B T)^{-1}\times$ the Helmholtz free energy if $\pi_x=e^{-E_x/k_B T}$ for some energy function $E_x$, Boltzmann constant $k_B$ and temperature $T$. Let the discrete gradient $\dgrad:\RR^\X\to\RR^\E$ be the adjoint of $-\ddiv$ from~\eqref{eq:disDiv}, i.e. $\dgrad_{xy} \xi\coloneqq\xi_{y}-\xi_x$, $\dgrad_{\In x} \xi\coloneqq\xi_{x}$. With this notation the quasipotential~\eqref{eq:QP} is related to the dynamic large deviations through the Hamilton-Jacobi-Bellman equation $\H(\rho,\dgrad\grad\V(\rho))=0$; this can be calculated explicitly but also follows abstractly from the large-deviation principle for the invariant measure, see for example \cite[Eq.~(2.7)]{BDSGJLL2002} or \cite[Thm.~3.6]{PRS2021TR}. Note 
 that here $\nabla \V(\rho)$ is a vector with elements $\partial V/\partial\rho_x$.

Without further assumptions, the quasipotential $\V$ is indeed a Lyapunov functional along the macroscopic dynamics~\eqref{eq:ODE}, which can be calculated explicitly 
\begin{equation}
\label{eq:free energy loss}
 \begin{aligned}
  &-\frac12\frac{d}{dt}\V\big(\rho(t)\big) 
  =
  \underbrace{\sumxny s\big( \rho_x Q_{xy} \mid \sqrt{\rho_x\rho_y\tfrac{\pi_x}{\pi_y}} Q_{xy}\big)
  +
  \sum_{x\in\X} \Bigl[  s\big(\lambda_{\In x} \mid \sqrt{\mfrac{\rho_x}{\pi_x}}\lambda_{\In x}\big)
  +
  s\big(\rho_x\lambda_{\Out x}\mid\sqrt{\rho_x\pi_x} \lambda_{\Out x}\big)\Bigr]}_{=\L^\asym(\rho,j^0(\rho))\geq0}\\
  & + \underbrace{\mfrac12\sumxny \Big( \sqrt{\rho_x Q_{xy}}-\sqrt{\rho_y Q_{xy}\mfrac{\pi_x}{\pi_y}}\Big)^2
  +
  \frac12\sum_{x\in\X} \Big\lbrack \big( \sqrt{\lambda_{\In x}}-\sqrt{\frac{\rho_x}{\pi_x}\lambda_{\In x}}\big)^2
        + \big( \sqrt{\rho_x\lambda_{\Out x}}-\sqrt{\pi_x\lambda_{\Out x}}\big)^2\Big\rbrack}_{=\Psi^*_{F^\asym}(\rho,F^\sym(\rho))\geq0}.
 \end{aligned}
\end{equation}
In Section~\ref{sec:decom} we introduce $\L^\asym(\rho,j^0(\rho))$ and see that it forms the cost of the macroscopic dynamics~\eqref{eq:hydroLim} if the underlying particle system is modified to a ``purely non-dissipative'' system; in the same section we introduce what we call the ``modified Fisher information'' $\Psi^*_{F^\asym}(\rho,F^\sym(\rho))$.

Before discussing the general setting, let us first discuss the detailed balance (equilibrium) case. The Markov process $\rho\super{n}(t)$ is in \emph{microscopic detailed balance} with respect to $\Pi\super{n}$ if the random path $t\mapsto(\rho\super{n}(t),W\super{n}(t))$ starting from $\rho\super{n}(0)\sim\Pi\super{n}$, $W\super{n}(0)=0$ has the same probability as the time-reversed path $t\mapsto(\rho\super{n}(T-t), W\super{n}(T)-W\super{n}(T-t))$ \cite[Sec.~4.1]{Renger2018a} \footnote{The time-reversed jump process~\cite{Norris1998} is analogous to the notion of time-reversed diffusion processes~\cite{Nelson1967}.} . For our simple setting, this notion of microscopic detailed balance is equivalent to what may be called \emph{macroscopic detailed balance}\footnote{
While microscopic detailed balance (with $Q$, $\pi$ replaced by $n$-particle counterparts $Q^{(n)}$, $\Pi^{(n)}$ in~\eqref{eq:DB}) is a condition on the flux of probability in the invariant measure $\Pi^{(n)}$ at finite-particle level, macroscopic detailed-balance is a condition about flux of mass in the macroscopic steady state $\pi$. For more involved systems, for instance chemical reaction networks, microscopic and macroscopic detailed balance need not be the same~\cite{AndersonCraciunKurtz2010}.}: 
\begin{align}\label{eq:DB}
  \pi_x Q_{xy} = \pi_y Q_{yx} &&\text{and}&& \lambda_{\In x} =  \pi_x\lambda_{\Out x}. 
\end{align}

On the large-deviation scale, (micro and macroscopic) detailed balance is equivalent to $\L(\rho,j)=\L(\rho,-j) + \langle \dgrad\grad\V(\rho),j\rangle$, which is in turn equivalent to $\H(\rho,\zeta)=\H(\rho,\dgrad\grad\V(\rho)-\zeta)$ by convex duality.

By contrast, if detailed balance does \emph{not} hold, then, starting from $\rho\super{n}(0),\rho\super{n}(T)\sim\Pi\super{n}(T)$, $W\super{n}(0)=0$, we obtain after time reversal that
$
    \overleftarrow\rho\super{n}(t)\coloneqq \rho\super{n}(T-t),
$
and
$
    \overleftarrow W\super{n}(t)\coloneqq W\super{n}(T)-W\super{n}(T-t)
$
are the normalised particle density and cumulative net flux of a \emph{different} particle system, where at each node $x$, new particles are created with rate $n\pi_x\lambda_{\Out x}$, each particle independently jumps to node $y$ with rate $\Q_{yx}\pi_{y}/\pi_{x}$ and is destroyed with rate $\lambda_{\In x}/\pi_x$, see again \cite[Sec.~4.1]{Renger2018a}. Analogous to \eqref{eq:LDP}, $(\overleftarrow \rho\super{n}(t),\overleftarrow W\super{n}(t))$ satisfies a large-deviation principle with rate functional $\int_0^T\!\overleftarrow{\L}\big(\rho(t),\dot w(t)\big)\,dt$, which is related to the original rate functional through the 
relation $\overleftarrow\L(\rho,j)=\L(\rho,-j) + \langle \dgrad\grad\V(\rho),j\rangle$, and by convex duality $\overleftarrow\H(\rho,\zeta)=\H(\rho,\dgrad\grad\V(\rho)-\zeta)$, see for example~\cite[Sec.~2.7]{BDSGJLL2002},\cite{MaesNetocny2003timereversal}, \cite[Sec.~4.2]{Renger2018a}.

\section{Force-dissipation decomposition and connections to Onsager-Machlup relation}
\label{sec:forces}

Our aim is now to decompose the large-deviation cost function~\eqref{def:Lag}
\begin{equation}\label{eq:gen-decom}
    \L(\rho,j) =\Psi(\rho,j)+\Psi^*\big(\rho,F(\rho)\big) - F(\rho)\cdot j,
\end{equation}
for some force field $F(\rho)\in\RR^\Edge$ and convex dual pair of non-negative \emph{dissipation potentials} $\Psi,\Psi^*$, i.e., for any $\rho$,
\begin{equation}\label{def:conv-dual}
    \Psi^*(\rho,\zeta)=\sup_{j} \zeta \cdot j - \Psi(\rho,j), \ \ \Psi(\rho,j)=\sup_{\zeta} \zeta \cdot j - \Psi^*(\rho,\zeta). 
\end{equation}
Let us first discuss the physical interpretation of decomposition~\eqref{eq:gen-decom}. As already hinted at above, of particular interest will be forces of the form $F(\rho)=-\frac12\dgrad \grad\V(\rho)$ in which case $\int_0^T\!F(\rho(t))\cdot j(t)\,dt=\frac12\int_0^T\!\grad\V(\rho(t))\cdot \dot \rho(t)\,dt=\frac12\V(\rho(T))-\frac12\V(\rho(0))$. This shows that the integrated version of \eqref{eq:gen-decom} has the dimension of entropy (or non-dimensionalised free energy), and so \eqref{eq:gen-decom} is really a power balance with its integrated version being a energy balance. For the zero-cost flow $\L(\rho,j)=0$, and so the sum $\Psi(\rho,j)+\Psi^*(\rho,F(\rho))$ models the dissipation of free energy or entropy, which justifies the term ``dissipation potentials''. In fact, by convex duality, the non-negativity of $\Psi,\Psi^*$ implies that $\Psi(\rho,0)\equiv0\equiv\Psi^*(\rho,0)$, reflecting the physical principle: \emph{there is no dissipation in the absence of fluxes and forces}. Therefore, $\Psi^*(\rho,F(\rho))=\L(\rho,0)$ is the energy that needs to be injected into the system in order to force $j=0$, and $\Psi(\rho,j)=\L(\rho,j)\rvert_{F(\rho)=0}$ is the energy that needs to be injected in order to force a nontrivial flux $j$ in the absence of forces.

Due to the duality~\eqref{def:conv-dual}, $\Psi$ and $\Psi^*$ are convex in their second argument and we have the inequality $\Psi^*(\rho,\zeta)+\Psi(\rho,j) \geq  \zeta\cdot j$ for any $j,\zeta$. Furthermore 
\begin{equation}
    \Psi^*(\rho,\zeta)+\Psi(\rho,j) =  \zeta\cdot j \ \Longleftrightarrow  \ \zeta =  \nabla_{j}\Psi(\rho,j) \ \Longleftrightarrow  \ j = \nabla_{\zeta}\Psi^*(\rho,\zeta).  
\end{equation}
When $\L=0$, which corresponds to the macroscopic flow $j^0$, the identity~\eqref{eq:gen-decom} along with the properties of convex duality imply that 
\begin{align}
    j^0(\rho) = \nabla_{\zeta} \Psi^*(\rho,F(\rho)) && \text{and} && F(\rho)=\nabla_{j} \Psi(\rho,j^0(\rho)).
\label{eq:full dynamics}
\end{align}
The first equality above provides a nonlinear response relation between forces and fluxes.

The decomposition~\eqref{eq:gen-decom} exists uniquely~\cite{MielkePeletierRenger14}, where the force and dual dissipation potential are explicitly given by \footnote{Strictly speaking~\cite{MielkePeletierRenger14} only covers the case of independent particles but the formulae follow naturally for systems with boundaries like the one discussed in this work.}
\begin{align}\label{eq:force}
    F_{xy}(\rho) &\coloneqq -\grad_{j_{xy}} \L(\rho,0) =
    \mfrac12 \log\mfrac{\rho_xQ_{xy}}{\rho_yQ_{yx}} 
    \quad\text{and}\quad 
    F_{\In x}(\rho)\coloneqq-\grad_{j_{\In x}}\L(\rho,0) = \mfrac12 \log\mfrac{\lambda_{\In x}}{\rho_x\lambda_{\Out x}},\\
    \Psi^*(\rho,\zeta) &\coloneqq \H\big(\rho,\zeta-F(\rho)\big)-\H\big(\rho,-F(\rho)\big) \notag\\
    &=
    2\sum_{\substack{x,y\in\X\\x<y}} \sqrt{\rho_x Q_{xy} \rho_y  Q_{yx}} \bigl( \cosh(\zeta_{xy}) -1 \bigr) + 2\sum_{x\in\X} \sqrt{\lambda_{\In x}\rho_x\lambda_{\Out x}}\bigl(\cosh(\zeta_{\In x})-1\bigr).
    \label{def:diss-pot}
\end{align}
The middle term
\begin{equation}
    \Psi^*(\rho,F(\rho))=\frac12\sumsum_{x\neq y}(\sqrt{\rho_xQ_{xy}}-\sqrt{\rho_yQ_{yx}})^2+\sum_x(\sqrt{\lambda_{\In x}}-\sqrt{\rho_x\lambda_{\Out x}})^2
\label{eq:Fisher}
\end{equation}
is often called the \emph{Fisher information}; it quantifies the energy needed to shut down all fluxes under force $F$, and also controls the long-time behaviour of the ergodic average $T^{-1}\int_0^T\rho(t)\,dt$ \cite{NuskenRenger2023}.

If detailed balance holds, the force is related to the quasipotential through $F(\rho)=-\frac12\dgrad\grad\V(\rho)$, which reflects the classical principle that systems in (macroscopic) detailed balance are completely driven by the free energy. 
This can be checked explicitly, but is also known to hold more generally~\cite{MielkePeletierRenger14}, since in that case the decomposition~\eqref{eq:gen-decom} can be interpreted as a generalised Onsager-Machlup relation. In particular, under detailed balance, the work done by the force along a trajectory equals the free-energy loss as
\begin{equation}
    F\big(\rho(t)\big)\cdot j(t)=-\mfrac12\dgrad\grad\V\big(\rho(t)\big)\cdot j(t)=\mfrac12\grad\V(\rho(t))\cdot\ddiv j(t)=-\mfrac12\grad\V(\rho(t))\cdot\dot\rho(t)=-\mfrac12\frac{d}{dt}\,\V(\rho(t)).
\label{eq:DB force}
\end{equation}

More generally without detailed balance, the cost function $\overleftarrow\L\!(\rho,j)$ of the time-reversed dynamics
admits a similar decomposition as in~\eqref{eq:gen-decom}, with the same dissipation potential~\eqref{def:diss-pot} and driving force $\overleftarrow F\!_{xy}(\rho) = \mfrac12 \log\mfrac{\rho_xQ_{yx}\pi_y/\pi_x}{\rho_yQ_{xy}\pi_x/\pi_y}$ and $\overleftarrow F\!_{\In x}(\rho)=\mfrac12 \log\mfrac{\pi_x\lambda_{\Out x}}{\rho_x\lambda_{\In x}/\pi_x}$.
This allows to 
define symmetric and antisymmetric forces with respect to time-reversal~\cite{BDSGJLL2015MFT,RengerZimmer2021,PRS2021TR}
\begin{align}\label{eq:Fsym Fasym}
     F^\sym_{xy}(\rho) &\coloneqq \mfrac12[F_{xy}(\rho)+\overleftarrow F\!_{xy}(\rho)]=
     \mfrac12\log\mfrac{\rho_x\pi_y}{\rho_y\pi_x}
     &&\text{and}&
     F^\sym_{\In x}(\rho) &\coloneqq\mfrac12[F_{\In x}(\rho)+\overleftarrow F\!_{\In x}(\rho)]=\mfrac12\log\mfrac{\pi_x}{\rho_x},\\
    F^\asym_{xy} &\coloneqq \mfrac12[F_{xy}(\rho)-\overleftarrow F\!_{xy}(\rho)]=
    \mfrac12\log\mfrac{\pi_x Q_{xy}}{\pi_y Q_{yx}}
    &&\text{and}& 
    F^\asym_{\In x} &\coloneqq\mfrac12[F_{\In x}(\rho)-\overleftarrow F\!_{\In x}(\rho)]=
    \mfrac12\log\mfrac{\lambda_{\In x}}{\pi_x\lambda_{\Out x}}.\notag
\end{align}
This decomposition of the force is natural and insightful because the symmetric force always takes the form $F^\sym(\rho)=-\frac12 \dgrad\nabla\V(\rho)$ as in the case of macroscopic detailed balance~\eqref{eq:DB} as explained above, and the antisymmetric force $F^\asym=0$ precisely if macroscopic detailed balance holds. So $F^\asym_{xy}$ and $F^\asym_{\In x}$ are exactly the bulk and boundary forces that drive the system out of detailed balance. This will be crucial to derive explicit expressions for the free energy loss~\eqref{eq:free energy loss} and the work done by the non-equilibrium force~\eqref{eq:antisym work} along the full dynamics~\eqref{eq:full dynamics}.

While it may seem surprising that $F^\asym$ is independent of $\rho$, it should be noted that this happens for various other systems as well~\cite[Sec.~5]{PRS2021TR}.

\section{Dissipative--non-dissipative decomposition of the cost}\label{sec:decom}

We will now use the notion of generalised orthogonality~\cite{KaiserJackZimmer2018,RengerZimmer2021,PRS2021TR} to further decompose the dual dissipation $\Psi^*$ in \eqref{eq:gen-decom} into purely dissipative and non-dissipative terms. 

To this end we first introduce the modified potential and the generalised pairing 
\begin{align}
    &\Psi^*_{\tilde\zeta}(\rho,\zeta)\coloneqq \label{eq:modified potential}\\
    &\qquad
    2\sumxly\sqrt{\rho_x Q_{xy} \rho_y Q_{yx}}\cosh(\tilde\zeta_{xy})\big(\cosh(\zeta_{xy})-1\big)
    + 2\sum_{x\in\X} \sqrt{\lambda_{\In x}\rho_x\lambda_{\Out x}} \cosh(\tilde\zeta_{\In x})\big(\cosh(\zeta_{\In x})-1\big),\notag\\
    &\theta_\rho(\zeta,\tilde\zeta)\coloneqq\notag\\
    &\qquad 2\sumxly\sqrt{\rho_x Q_{xy} \rho_y Q_{yx}}\sinh(\tilde\zeta_{xy})\sinh(\zeta_{xy})
    + 2\sum_{x\in\X} \sqrt{\lambda_{\In x}\rho_x\lambda_{\Out x}} \sinh(\tilde\zeta_{\In x})\sinh(\zeta_{\In x}).\notag
\end{align}
Using the addition rule $\cosh(\zeta+\tilde\zeta)=\cosh(\zeta)\cosh(\tilde\zeta)+\sinh(\zeta)\sinh(\tilde\zeta)$, one finds that dual dissipation $\Psi^*$~\eqref{def:diss-pot} can be expanded as
$\Psi^*(\rho,\zeta+\tilde\zeta)= \Psi^*(\rho,\tilde\zeta) + \theta_\rho(\zeta,\tilde\zeta) + \Psi^*_{\tilde\zeta}(\rho,\zeta)$.

Of particular interest is the case where $\zeta=F^\sym(\rho),\tilde\zeta=F^\asym$. Using the explicit expression for the forces \eqref{eq:Fsym Fasym} and the definition of $\sinh$ in terms of exponential function we find
\begin{align*}
    &\theta_\rho\big(F^\sym(\rho),F^\asym\big)\\
    &\qquad=4\sum_{\substack{x,y\in\X\\x<y}} \sqrt{\rho_x Q_{xy} \rho_y  Q_{yx}} \sinh(F^\sym_{xy})\sinh(F^\asym_{xy})
    +4\sum_{x\in\X} \sqrt{\lambda_{\In x}\rho_x\lambda_{\Out x}}\sinh(F^\sym_{\In x})\sinh(F^\asym_{\In x})\\
    &\qquad=\sum_{x\in\X} \mfrac{\rho_x}{\pi_x} \underbrace{\Big\lbrack\sum_{\substack{y\in\X,\\ y\neq x}} (\pi_x Q_{xy} - \pi_y Q_{yx}) 
    + \pi_x\lambda_{\Out x}-\lambda_{\In x}\Big\rbrack}_{=0}
    + \underbrace{\sum_{x\in\X}( \lambda_{\In x}-\pi_x\lambda_{\Out x})}_{=0}=0.
\end{align*}
The fact that the generalised cross term $\theta\big(F^\sym,F^\asym\big)$ vanishes reflects an orthogonality of the symmetric and antisymmetric forces in a generalised sense~\cite[Prop.~4.2]{RengerZimmer2021}, \cite[Prop.~2.24]{PRS2021TR}.   

This orthogonality is also related to the quasipotential as follows. First, consider a system with free energy $\V$ and force $F=F^\sym+F^\asym, F^\sym=-\frac12\dgrad\grad\V$. Then $\V$ is also the quasipotential for the modified system where the nondissipative force $F^\asym$ is replaced by zero, i.e.
\begin{equation*}
    \H^\sym\big(\rho,\dgrad\grad\V(\rho)\big)=\theta_\rho\big(-F^\sym(\rho),0\big)=0.
\end{equation*}
Second, consider a system in detailed balance with quasipotential $\V$ and $F=F^\sym=-\frac12\dgrad\grad\V$. If one would add an additional force $\zeta$, the modified Hamilton-Jacobi equation reads
\begin{equation*}
    \begin{aligned}
    \H^{\sym,\zeta}\big(\rho,\dgrad\grad\V(\rho)\big)&\coloneqq \Psi^*\big(\rho,\dgrad\grad\V(\rho)+F^\sym+\zeta\big)-\Psi^*\big(\rho,F^\sym+\zeta\big)\\
    &=
    \Psi^*\big(\rho,-F^\sym+\zeta\big)-\Psi^*\big(\rho,F^\sym+\zeta\big)=-2\theta_\rho(F^\sym,\zeta).
    \end{aligned}
\end{equation*}
Thus, the forces $\zeta$ orthogonal to $F^\sym$ are precisely those forces that leave the quasipotential invariant when added to a symmetric force. Therefore this orthogonality means that the quasipotential $\V$ and steady state $\pi$ are unaltered by turning $F^\asym$ on or off, which is also observed in the numerical examples in Section~\ref{sec:numerics}.

Therefore we can expand $\Psi^*$ to arrive at following splitting of the Fisher information
\begin{equation*}
\begin{aligned}
    \Psi^*\big(\rho,F(\rho)\big)=\Psi^*\big(\rho,F^\sym(\rho)+F^\asym\big)
    &=\Psi^*\big(\rho,F^\asym\big) 
    + \underbrace{\theta_\rho\big(F^\sym(\rho),F^\asym\big)}_{=0}
    + \Psi^*_{F^\asym}\big(\rho,F^\sym(\rho)\big)\\
    &=\Psi^*\big(\rho,F^\sym(\rho)\big)  
       +\underbrace{\theta_\rho\big(F^\sym(\rho),F^\asym\big)}_{=0}  
    +\Psi^*_{F^\sym}\big(\rho,F^\asym\big).
\end{aligned}
\end{equation*}
These expansions are not quite the same as in the quadratic case -- one of the potentials needs to be modified according to \eqref{eq:modified potential}. This yields the \emph{modified Fisher informations}, cf.\ \eqref{eq:Fisher},
\begin{align*}
    &\Psi^*_{F^\asym}\big(\rho,F^\sym(\rho)\big)\\
    &\quad=\mfrac12\sumxny \big(\sqrt{\rho_x Q_{xy}}-\sqrt{\rho_y\mfrac{\pi_x}{\pi_y}Q_{xy}}\big)^2
    +
    \mfrac12\sum_{x\in\X} \big( \sqrt{\lambda_{\In x}}-\sqrt{\frac{\rho_x}{\pi_x}\lambda_{\In x}}\big)^2 
    +
    \mfrac12\sum_{x\in\X} \big( \sqrt{\rho_x\lambda_{\Out x}}-\sqrt{\pi_x\lambda_{\Out x}}\big)^2,\\
    &\Psi^*_{F^\sym}(\rho,F^\asym) \notag\\ 
    &\quad=
    \mfrac12\sumxny\big( \sqrt{\rho_x Q_{xy}}-\sqrt{\rho_x \mfrac{\pi_y}{\pi_x}Q_{yx}}\big)^2
    +
    \mfrac12\sum_{x\in\X} \big(\sqrt{\lambda_{\In x}}-\sqrt{\pi_x\lambda_{\Out x}}\big)^2
    +
    \mfrac12\sum_{x\in\X} \big(\sqrt{\rho_x \mfrac{\lambda_{\In x}}{\pi_x}} - \sqrt{\rho_x \lambda_{\Out x}}\big)^2.
\end{align*}

Applying this expansion of dissipation potentials to~\eqref{eq:gen-decom} leads to two distinct and physically relevant  decompositions
\begin{subequations}
\begin{align}
    \L(\rho,j)  &= \underbrace{\Psi(\rho,j)+\Psi^*(\rho,F^\asym)-F^\asym\cdot j}_{\eqqcolon\L^\asym(\rho,j)} + \Psi^*_{F^\asym}\big(\rho,F^\sym(\rho)\big) - F^\sym(\rho)\cdot j,\label{eq:L-Fsym-split}\\
    &= \underbrace{\Psi(\rho,j)+\Psi^*\big(\rho,F^\sym(\rho)\big)-F^\sym\cdot j}_{\eqqcolon\L^\sym(\rho,j)} + \Psi^*_{F^\sym}(\rho,F^\asym) - F^\asym\cdot j.
    \label{eq:L-Fasym-split}
\end{align}
\label{eq:L-split}
\end{subequations}
The two ``modified cost functions'' $\L^\sym,\L^\asym$ are non-negative by convex duality, and are in fact themselves large-deviation cost functions of particle system with modified jump rates, see Appendix~\ref{app:modified cost functions}. Since $F^\sym=-\frac12\dgrad\grad\V$, the symmetric cost $\L^\sym$ encodes the (non-quadratic) Onsager-Machlup dissipative (gradient-flow) part of the dynamics, even without assuming detailed balance. By analogy, $\L^\asym$ encodes a \emph{non-dissipative} dynamics that is in some sense the time-antisymmetric counterpart of a gradient flow; this will be explored in Section~\ref{sec:Ham}. Both expressions~\eqref{eq:L-split} decompose the cost function $\L$ into terms corresponding to the dissipative and non-dissipative dynamics, but because $\Psi^*$ is non-quadratic, there are two distinct ways to do so\footnote{\eqref{eq:L-Fsym-split} is related to ``FIR inequalities'' that have been used to study singular limits and prove errors estimates~\cite{DuongLamaczPeletierSharma17,DLPSS17,PeletierRenger21}; the cost $\L^\asym$ quantifies the gap in the inequality.}. 


Of particular interest are the decompositions~\eqref{eq:L-split} along the zero-cost flux $j^0(\rho)$. The work done by the symmetric force is $F^\sym\cdot j^0 = -\frac12\frac{d}{dt}\dgrad\nabla\V$, so that we retrieve the free-energy loss~\eqref{eq:free energy loss} from~\eqref{eq:L-Fsym-split}, with the explicit expression for $\L^\asym(\rho,j^0)$ given by the $s(\cdot|\cdot)$ terms in~\eqref{eq:free energy loss}. Analogously, inserting $j^0$ into~\eqref{eq:L-Fasym-split}, we find an explicit expression for the work done by the antisymmetric force
\begin{equation}
    \int_0^T F^\asym\cdot j^0\big(\rho(t)\big)\,dt =  - \int_0^T\Bigl[\L^\sym\bigl(\rho(t),j^0(t)\bigr) + \Psi_{F^\asym}^*\bigl(\rho(t),F^\sym(\rho(t))\bigr) \Bigr]\,dt \leq0
\label{eq:antisym work}    
\end{equation}
with 
\begin{align}\label{eq:Lsym-j0}
\begin{aligned}
    &\L^\sym\big(\rho, j^0(\rho)\big)=\\
     &\sumxny s\big( \rho_x Q_{xy} \mid \rho_x\sqrt{\tfrac{\pi_x}{\pi_y}Q_{xy}Q_{yx}} \big) 
    + 
    \sum_{x\in\X}\big\lbrack 
        s\big( \lambda_{\In x} \mid \sqrt{\pi_x\lambda_{\In x}\lambda_{\Out x}}  \big) + s\big( \rho_x\lambda_{\Out x} \mid \rho_x  \sqrt{\tfrac{\lambda_{\In x}\lambda_{\Out x}}{\pi_x}} \big)
    \bigr]. 
\end{aligned}
\end{align}
While, a priori, both $\L^\sym(\rho,j)$ and $\L^\asym(\rho,j)$ appear as a minimisation over one-way fluxes as in~\eqref{def:Lag} (see Appendix~\ref{app:modified cost functions}), for $j=j^0(\rho)$, the minimising one-way flux is exactly $j^+_{xy}=\rho_x Q_{xy}$, $j^+_{\In x}=\lambda_{\In x}$ which considerably simplifies the expressions. 

As mentioned in the introduction, the non-positivity of the antisymmetric work \eqref{eq:antisym work} was derived for chemical reactions in \cite[eq.~(18)]{FreitasEsposito2022}.

\section{Dissipative and non-dissipative zero-cost dynamics}\label{sec:Ham}

Recall from Section~\eqref{sec:forces} that $\L=0$ for the full macroscopic dynamics and so $\dot\rho=-\ddiv\grad_\zeta\Psi^*(\rho,F^\sym(\rho)+F^\asym)$. Similarly $\L^\sym=0$ yields the nonlinear \emph{gradient flow} $\dot\rho=-\ddiv\grad_\zeta\Psi^*(\rho,-\frac12\dgrad\grad\V(\rho))$ driven by the free energy $\V$. How can the zero-cost dynamics of $\L^\asym$ be given a physical interpretation? The ODE describing this dynamics is
\begin{align}
    \dot\rho_x(t) &= -\ddiv_x \grad_\zeta\Psi^*(\rho,F^\asym)\notag\\
    &=
    \sumsum_{\substack{x,y\in\X\\x\neq y}} \sqrt{\rho_x(t)\rho_y(t)} \Bigl( Q_{yx}\sqrt{\tfrac{\pi_y}{\pi_x}}-Q_{xy}\sqrt{\tfrac{\pi_x}{\pi_y}}\Bigr) +\sqrt{\rho_x} \Bigl( \lambda_{\In x}\tfrac{1}{\sqrt{\pi_x}} - \lambda_{\Out x}\sqrt{\pi_x} \Bigr).
\label{eq:asym ODE}
\end{align}
Our novel and maybe surprising result
is that this equation in fact has a Hamiltonian structure $\dot\rho=\JJ(\rho)\grad\U(\rho)$ with energy and Poisson structure given by
\begin{align}
  \U(\rho)=&\sum_{x\in\X}(\sqrt{\pi_x}-\sqrt{\rho_x})^2, 
\label{eq:Hamiltonian energy}\\
  \JJ_{xy}(\rho)=&2\sum_{z\in\X} \sqrt{\rho_x\rho_y\rho_z}
    \Big\lbrack
      \sqrt{\mfrac{\pi_x\pi_z}{\pi_y}}Q_{zy}-\sqrt{\mfrac{\pi_x\pi_y}{\pi_z}}Q_{yz} - \sqrt{\mfrac{\pi_y\pi_z}{\pi_x}}Q_{zx} + \sqrt{\mfrac{\pi_x\pi_y}{\pi_z}}Q_{xz}
    \Big\rbrack\notag\\
    &+
    2\sqrt{\rho_x\rho_y}
    \Big\lbrack
      \sqrt{\mfrac{\pi_x}{\pi_y}}\lambda_{\In y} - \sqrt{\pi_x\pi_y}\lambda_{\Out y} - \sqrt{\mfrac{\pi_y}{\pi_x}}\lambda_{\In x} + \sqrt{\pi_x\pi_y}\lambda_{\Out x}  
    \Big\rbrack,
    \qquad
  x,y\in \X, x\neq y.
\label{eq:Poisson structure}
\end{align}
We include a brief derivation in Appendix~\ref{app:Hamilton} and in Appendix~\ref{app:Jacobi} verify that the corresponding Poisson bracket $\lbrack \mathscr F^1,\mathscr F^2\rbrack_\rho\coloneqq\grad\mathscr F^1(\rho)\cdot\JJ(\rho)\grad\mathscr F^2(\rho)$ satisfies the Jacobi identity (requisite for a Hamiltonian system). The energy \eqref{eq:Hamiltonian energy} is known as the Hellinger distance~\cite{Hellinger1909}, mostly used in statistics~\cite{Beran1977} and recently also to describe certain reaction dynamics as gradient flows~\cite{LieroMielkeSavare2018}.

The Hamiltonian structure $(\U,\JJ)$ for the ODE~\eqref{eq:asym ODE} is generally not unique. In contrast to the gradient flow for $\L^\sym$, it is not clear to us whether $\U$ and $\JJ$ are somehow related to the variational structure provided by $\L^\asym$. A natural question is then whether -- in the spirit of metriplectic systems~\cite{Morrison1986} or GENERIC~\cite{Ottinger2005} -- there could be a Hamiltonian structure for \eqref{eq:asym ODE} so that the energy $\U$ is also conserved along the full dynamics $\L=0$. The answer to this question is negative, because by~\eqref{eq:free energy loss} the full dynamics simultaneously dissipates free energy until the unique steady state is reached. Another fundamental difference with GENERIC is that here the full dynamics is retrieved by adding the forces $F=F^\sym+F^\asym$, whereas in GENERIC one retrieves the full dynamics by adding velocities or fluxes.

\section{Insights from a simple system}
\label{sec:numerics}

Consider the simple example of Figure~\ref{fig:open-bound-chain} with $\X=\{A,B,C\}$ and define the positive edges as $\E=\{(A,B),(A,C),(B,C)\}\cup\{\In A, \In C\}$ (with no in/out-flow at $B$). In what follows we use $j^0, j^{\sym,0}, j^{\asym,0}$ for the zero-cost flux for $\L,\L^\sym,\L^\asym$ respectively.  

\begin{figure}[ht]
        \centering
         \begin{subfigure}{0.32\textwidth}
          \centering
          \includegraphics[width=\textwidth]{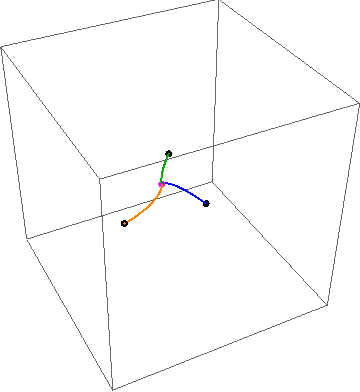}
        \end{subfigure}
        \begin{subfigure}{0.32\textwidth}
                \centering
                \includegraphics[width=\textwidth]{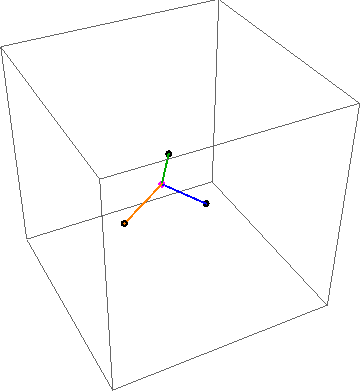}
        \end{subfigure}
         \begin{subfigure}{0.32\textwidth}
            \centering
            \includegraphics[width=\textwidth]{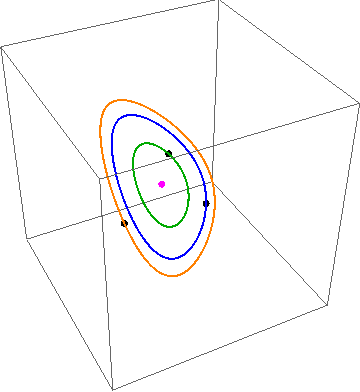}
        \end{subfigure}
         \begin{subfigure}{0.32\textwidth}
            \centering
            \begin{tikzpicture}[>=stealth, node distance=7em]
            \tikzstyle{every node}=[font=\scriptsize];
            \node[state, fill=antiquewhite] (B) {{\color{magenta}$\pi_B=\frac13$}};
            \node[state, below right of=B, fill=antiquewhite] (C) {{\color{magenta}$\pi_C=\frac13$}};
            \node[state, below left of=B, fill=antiquewhite] (A) {{\color{magenta}$\pi_A=\frac13$}};
            
            \draw (B) edge[->] node[above,pos=0.5,sloped] {$\frac13$} (C);
            \draw (A) edge[->] node[above,pos=0.5,sloped] {$\frac13$} (B);
            \draw (C) edge[->] node[below,pos=0.5,sloped] {$\frac13$} (A);
            \end{tikzpicture}
        \end{subfigure}
        \begin{subfigure}{0.32\textwidth}
            \centering
            \begin{tikzpicture}[>=stealth, node distance=7em]
            \tikzstyle{every node}=[font=\scriptsize];
            \node[state, fill=antiquewhite] (B) {{\color{magenta}$\pi_B=\frac13$}};
            \node[state, below right of=B, fill=antiquewhite] (C) {{\color{magenta}$\pi_C=\frac13$}};
            \node[state, below left of=B, fill=antiquewhite] (A) {{\color{magenta}$\pi_A=\frac13$}};
            
            \draw (B) edge[-] node[above,pos=0.5,sloped] {$0$} (C);
            \draw (A) edge[-]  node[above,pos=0.5,sloped] {$0$} (B);
            \draw (C) edge[-] node[below,pos=0.5,sloped] {$0$} (A);
            \end{tikzpicture}            
        \end{subfigure}
         \begin{subfigure}{0.32\textwidth}
            \centering
            \begin{tikzpicture}[>=stealth, node distance=7em]
            \tikzstyle{every node}=[font=\scriptsize];
            \node[state, fill=antiquewhite] (B) {{\color{magenta}$\pi_B=\frac13$}};
            \node[state, below right of=B, fill=antiquewhite] (C) {{\color{magenta}$\pi_C=\frac13$}};
            \node[state, below left of=B, fill=antiquewhite] (A) {{\color{magenta}$\pi_A=\frac13$}};
            
            \draw (B) edge[->] node[above,pos=0.5,sloped] {$\frac13$} (C);
            \draw (A) edge[->]  node[above,pos=0.5,sloped] {$\frac13$} (B);
            \draw (C) edge[->] node[below,pos=0.5,sloped] {$\frac13$} (A);
            
            \end{tikzpicture}                        
        \end{subfigure}
       \caption{Case A: Pure bulk effects. Top row: Plots of the zero-cost trajectories $\rho(t)$ associated to $j^0$, $j^{0,\sym}$, $ j^{0,\asym}$, starting from three different initial conditions (black dots) with the steady states depicted by the pink dots. 
       Bottom row: The steady states $\pi$ (in pink) and steady-state fluxes (magnitude indicated by values and direction by arrows) corresponding to $j^0$, $j^{0,\sym}$, $j^{0,\asym}$ respectively.}
\label{fig:A 0-traj}
\end{figure}
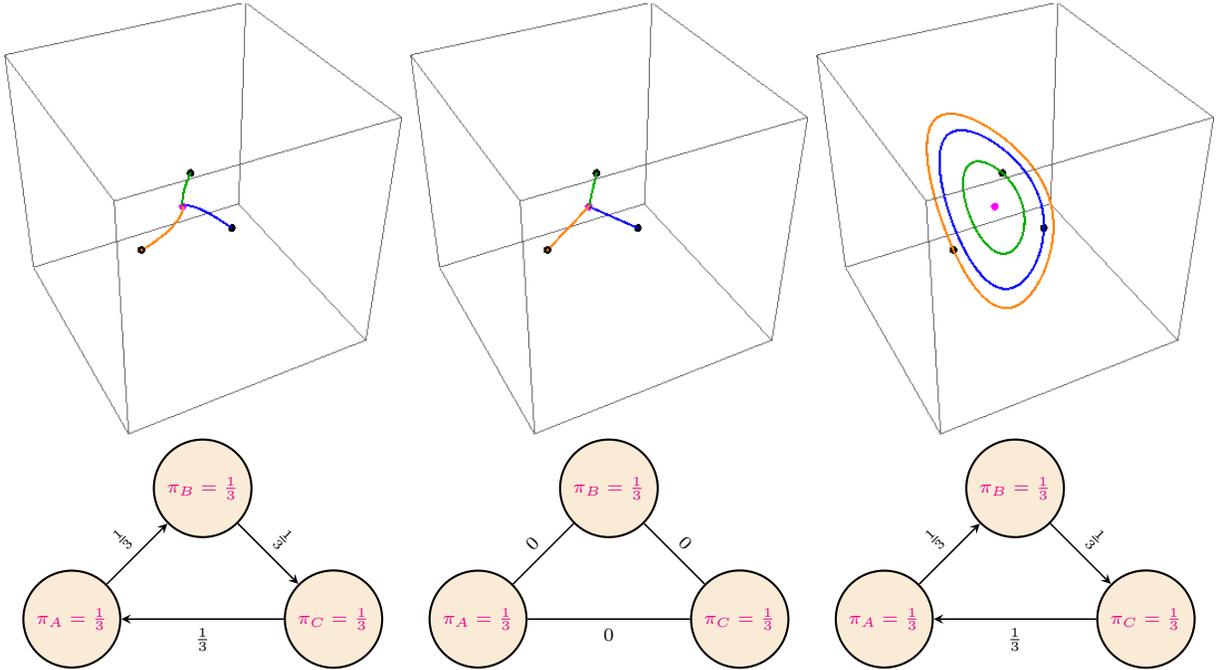

\textbf{Case A: Pure bulk effects.} We assume that the forward transition rates $Q_{AB}=Q_{BC}=Q_{CA}=2$ and backward transition rates $Q_{AB}=Q_{BC}=Q_{CA}=1$ and $\lambda_{\In x}=\lambda_{\Out x} = 0$ for $x=A,B$. This corresponds to a closed system being driven out of detailed balance purely by the bulk force, which is encoded in the different forward and backward transition rates (no detailed balance). Since there is no in and outflow, the total mass of the system is preserved at all times (and equal to the mass at $t=0$) with the steady state $\pi=(\frac13,\frac13,\frac13)$. The zero-cost trajectories and corresponding steady states are plotted in Figure~\ref{fig:A 0-traj}.

There are three interesting observations about the trajectories. First, in line with preceding discussions, both the full and symmetric zero-cost trajectories (top row, left \& middle) converge to the steady state $\pi$ whereas the antisymmetric zero-cost trajectory (top row, right) orbits around the steady state. Second, that all the trajectories are confined to a plane which corresponds to the conservation of total mass ($\sum_x \rho_x(t)=1$). Third, the symmetric zero-cost trajectories are straight lines since the purely dissipative dynamics is a gradient flow of a linear system (since there is no in/out flow).

From the steady states we see that, as expected, the symmetric zero-cost dynamics has an equilibrium/detailed balanced steady state (bottom row, middle), and the full system (bottom row, left) has a non-equilibrium steady state. Surprisingly, the (static) steady state $\pi$ of the antisymmetric dynamics (bottom row, right) leaves the steady state and even the corresponding flux of the full system unchanged. This is in line with the observation that the forces orthogonal to the symmetric force are precisely the ones that leave the quasipotential unchanged (see Section~\ref{sec:decom}).

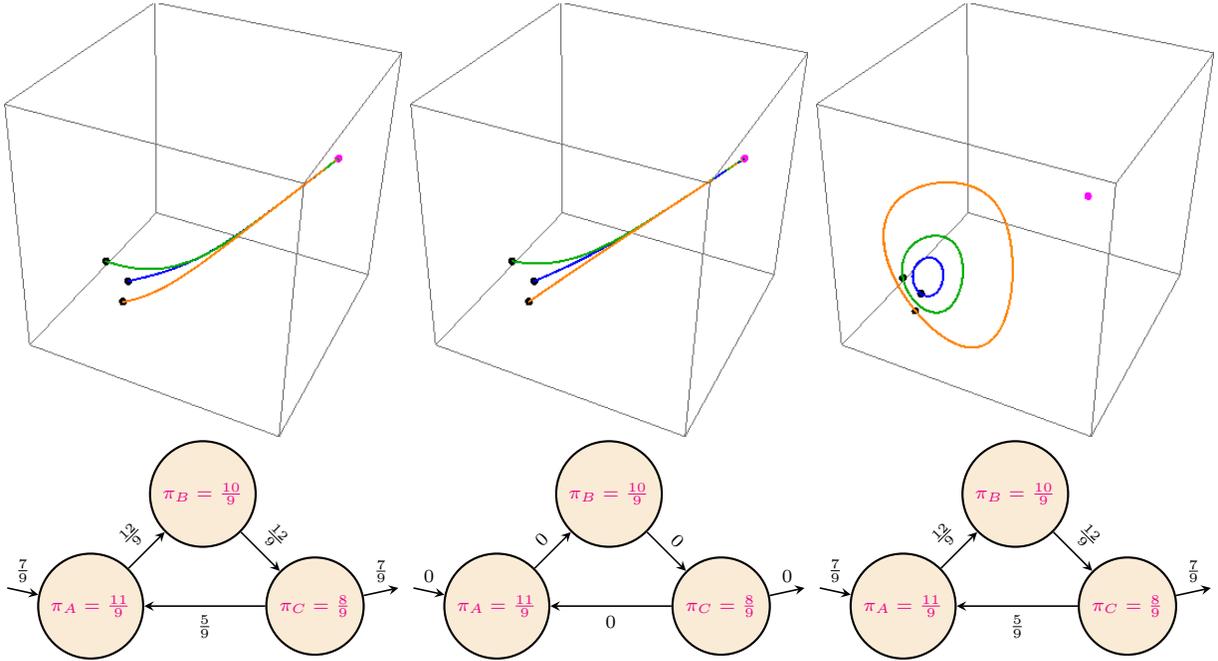
\begin{figure}[ht]
        \centering
        \begin{subfigure}{0.32\textwidth}
          \centering
          \includegraphics[width=\textwidth]{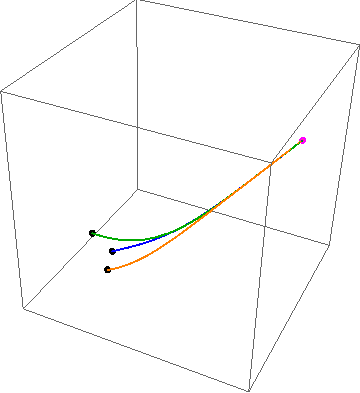}
        \end{subfigure}
        \begin{subfigure}{0.32\textwidth}
                \centering
                \includegraphics[width=\textwidth]{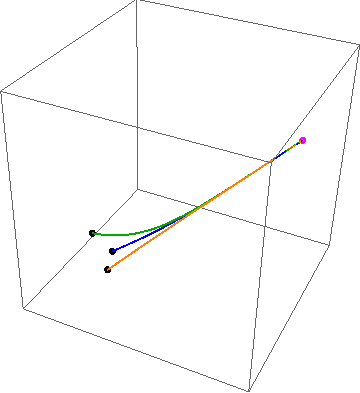}
        \end{subfigure}
         \begin{subfigure}{0.32\textwidth}
                \centering
                \includegraphics[width=\textwidth]{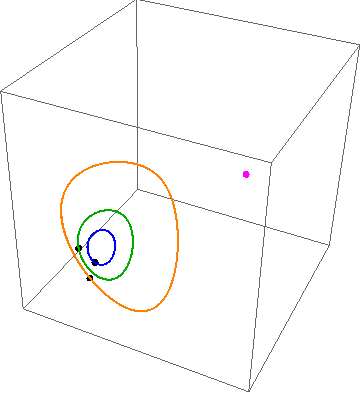}
        \end{subfigure}
         \begin{subfigure}{0.32\textwidth}
            \centering
            \begin{tikzpicture}[>=stealth,thick,scale=0.5, node distance=6em] 
            \tikzstyle{every node}=[font=\scriptsize];
            \node[state, fill=antiquewhite] (B) {{\color{magenta}$\pi_B=\frac{10}{9}$}};
            \node[state, below right of=B, fill=antiquewhite] (C) {{\color{magenta}$\pi_C=\frac{8}{9}$}};
            \node[state, below left of=B, fill=antiquewhite] (A) {{\color{magenta}$\pi_A=\frac{11}{9}$}};
            
            \draw (B) edge[->] node[above,pos=0.5,sloped] {$\frac{12}{9}$} (C);
            \draw (A) edge[->] node[above,pos=0.5,sloped] {$\frac{12}{9}$} (B);
            \draw (C) edge[->] node[below,pos=0.5,sloped] {$\frac{5}{9}$} (A);
            
            \draw[semithick,->] (C) to  [left=14] node[above] {$\frac{7}{9}$} (5.2,-2.5) ;
            \draw[semithick,<-] (A) to [right=14] node[above] {$\frac{7}{9}$} (-5.2,-2.5);
            \end{tikzpicture}
        \end{subfigure}
        \begin{subfigure}{0.32\textwidth}
            \centering
            \begin{tikzpicture}[>=stealth,thick,scale=0.5, node distance=6em]
            \tikzstyle{every node}=[font=\scriptsize];
            \node[state, fill=antiquewhite] (B) {{\color{magenta}$\pi_B=\frac{10}{9}$}};
            \node[state, below right of=B, fill=antiquewhite] (C) {{\color{magenta}$\pi_C=\frac{8}{9}$}};
            \node[state, below left of=B, fill=antiquewhite] (A) {{\color{magenta}$\pi_A=\frac{11}{9}$}};
            
            \draw (B) edge[->] node[above,pos=0.5,sloped] {$0$} (C);
            \draw (A) edge[->] node[above,pos=0.5,sloped] {$0$} (B);
            \draw (C) edge[->] node[below,pos=0.5,sloped] {$0$} (A);
            
            \draw[semithick,->] (C) to  [left=14] node[above] {$0$} (5.2,-2.5) ;
            \draw[semithick,<-] (A) to [right=14] node[above] {$0$} (-5.2,-2.5);
            \end{tikzpicture}
        \end{subfigure}
         \begin{subfigure}{0.32\textwidth}
            \centering
            \begin{tikzpicture}[>=stealth,thick,scale=0.5, node distance=6em] 
            \tikzstyle{every node}=[font=\scriptsize];
            \node[state, fill=antiquewhite] (B) {{\color{magenta}$\pi_B=\frac{10}{9}$}};
            \node[state, below right of=B, fill=antiquewhite] (C) {{\color{magenta}$\pi_C=\frac{8}{9}$}};
            \node[state, below left of=B, fill=antiquewhite] (A) {{\color{magenta}$\pi_A=\frac{11}{9}$}};
            
            \draw (B) edge[->] node[above,pos=0.5,sloped] {$\frac{12}{9}$} (C);
            \draw (A) edge[->] node[above,pos=0.5,sloped] {$\frac{12}{9}$} (B);
            \draw (C) edge[->] node[below,pos=0.5,sloped] {$\frac{5}{9}$} (A);
            
            \draw[semithick,->] (C) to  [left=14] node[above] {$\frac{7}{9}$} (5.2,-2.5) ;
            \draw[semithick,<-] (A) to [right=14] node[above] {$\frac{7}{9}$} (-5.2,-2.5);
            \end{tikzpicture}
        \end{subfigure}
        \caption{Case B: bulk and boundary effects. Top row: Plots of the zero-cost trajectories $\rho(t)$ associated to $j^0$, $j^{0,\sym}$, $ j^{0,\asym}$,  starting from three initial conditions (black dotes) with the steady states denotes by the pink dots. The initial conditions are the same as in Figure~\ref{fig:A 0-traj}. Bottom row: The steady states $\pi$ (in pink) and steady-state fluxes (magnitude indicated by values and direction by arrows) corresponding to to $j^0$, $j^{0,\sym}$, $ j^{0,\asym}$ respectively.}
\label{fig:B 0-traj}
\end{figure}
\textbf{Case B: Bulk and boundary effects.} 
As in Case A we assume that $Q_{AB}=Q_{BC}=Q_{CA}=2$ and $Q_{AB}=Q_{BC}=Q_{CA}=1$. For the boundary we assume that $\lambda_{\In A}=\lambda_{\Out C} = 2$ and $\lambda_{\Out A}=\lambda_{\In C} = 1$. This case corresponds to the system being driven out of detailed balance by both bulk and boundary effects. Regardless of initial condition, the steady state $\pi=(\frac{11}{9},\frac{10}{9},\frac{8}{9})$ is unique and positive but no longer a probability density, see Appendix~\ref{app:steady-state}. The zero-cost trajectories and corresponding steady states are plotted in Figure~\ref{fig:B 0-traj}.

As in the previous case, both the full and symmetric zero-cost trajectories (top row, left \& middle) converge to the steady state $\pi$ while the antisymmetric zero-cost trajectory orbits around the static steady state (top row, right), however, with the crucial difference that the trajectories are no longer confined to a plane since the mass is not conserved due to in/out flow at the nodes. We point out that the trajectories of the full and symmetric system are different even though they appear to be the quite close from the figures (compare in particular the orange trajectory in the top row of Figure~\ref{fig:B 0-traj}). 

A natural next step is to study the behaviour of the system under varying combinations of symmetric and antisymmetric bulk and boundary forces. Consider for example the system of case B, where the force is replaced by $\tilde F_{xy}\coloneqq F^\sym_{xy}$ and $\tilde F_{\In x}\coloneqq F^\asym_{\In x}$, i.e.\ purely symmetric bulk force and antisymmetric boundary force. This altered system will also have an altered steady state $\tilde\pi$, and as a consequence, the decomposition into symmetric and antisymmetric forces will be different, that is, in general $\tilde F^\sym\neq F^\sym, \tilde F^\asym\neq F^\asym$. 
In fact, it is impossible to construct a system where the bulk is in detailed balance ($F_{xy}=F_{xy}^\sym$) but the boundaries are not ($F_{\In x}\neq F^\sym_{\In x}$). Indeed, the steady state corresponding to such system would have some nodes with non-trivial in and outflow, but since the bulk has zero net fluxes, mass cannot be transported from the inflow to the outflow nodes. 
By contrast, take the system of case A with the family of uniform steady states $\pi=(a,a,a), a>0$. If one now adds boundary forces such that $\lambda_{\In x}/\lambda_{\Out x}=\pi_x=a$ for some $a>0$, then the steady state of the altered system is still $\pi=(a,a,a)$. One can thus construct a system where the bulk is not in detailed balance ($F_{xy}\neq F_{xy}^\sym$) but the boundaries are ($F_{\In x}=F^\sym_{\In x}$).

\section{Discussion}

As pioneered by Onsager and Machlup, microscopic fluctuations on the large-deviation scale provide a free energy balance for the macroscopic dynamics. By taking fluxes into account, macroscopic fluctuation theory extends this principle to non-equilibrium systems to obtain explicit balances \eqref{eq:gen-decom}, \eqref{eq:free energy loss} and \eqref{eq:antisym work} in terms of the work done by the full, symmetric and antisymmetric forces $F, F^\sym, F^\asym$ respectively.

With the aim of understanding the role of bulk and boundary effects in non-equilibrium non-diffusive systems, we study an open linear system on a graph. The derivation of the three energy balances poses a number of challenges. First, we derive the explicit quasipotential \eqref{eq:QP} (free energy) as the large-deviation rate of the microscopic invariant measure. Second, since the microscopic fluctuations are Poissonian rather than white noise, the large-deviation cost $\L$ cost is non-quadratic and therefore requires a generalised notion of orthogonality of forces. Whereas the modified system $\L^\sym=0$ is purely driven by the dissipation of free energy, the third challenge is to understand the system $\L^\asym=0$. As observed for closed linear systems in~\cite{PRS2021TR}, it turns out that with open boundaries, this dynamics is indeed a Hamiltonian system -- even satisfying the Jacobi identity. Our work thus allows to distinguish between dissipative (symmetric) and non-dissipative (antisymmetric/Hamiltonian) boundary and bulk mechanisms. We expect that these ideas will apply to more general nonlinear networks, for instance open networks with zero-range interactions (and related agent-based models in social sciences) and chemical-reaction networks attached to reservoirs. 

A few intriguing questions emerge from our analysis in regards to the role of antisymmetric forces. It turns out the antisymmetric forces are the exactly the ones that leave the quasipotential and steady state invariant (Section~\ref{sec:decom}). This leads to the natural question if one can optimise these forces in a systematic manner to speed up convergence to equilibrium; this is an important challenge in sampling of free energy in computational statistical mechanics~\cite{LelievreNierPavliotis13,BelletSpiliopoulos15,BelletSpiliopoulos16,DuncanLelievrePavliotis16,KaiserJackZimmer17}. Finally, it may be intuitively clear that the antisymmetric flow, as the opposite of a dissipative dynamics, should be non-dissipative, the appearance of a full Hamiltonian system with the Hellinger distance as conserved energy seems rather surprising and it is not well understood how and why this structure emerges. So far we have only managed to prove the Jacobi identity for systems where the weights (or jump rates) encoded in $Q$ are constant, while we do observe periodic orbits and prove conservation of an energy more generally. It is not at all clear if the Jacobi identity is only a feature of jump systems with constant rates or satisfied more generally.

\noindent\paragraph{Acknowledgements.} We thank the anonymous referees for their comments which help improve this manuscript considerably. The work of MR has been funded by Deutsche Forschungsgemeinschaft (DFG) through grant CRC 1114 ``Scaling Cascades in Complex Systems'' Project C08, Project Number 235221301. The work of US is supported by the Alexander von Humboldt foundation.

\appendix
\section{Appendix}

\subsection{Invariant measure and steady state}\label{app:steady-state}

\paragraph{Product-Poisson form of $\Pi\super{n}$.}
We show that the invariant measure $\Pi\super{n}$ for the (underlying) random process $\rho\super{n}(t)$ (described in Section~\ref{sec:model}) indeed has the explicit expression~\eqref{eq:product-Poisson}, i.e.\ it satisfies the backward equation
\begin{equation}\label{eq:abstract-stat}
    \sum_{\rho\in (\frac1n\NN_0)^\X} \Pi\super{n}(\rho)(\Q\super{n} f)(\rho)=0,
\end{equation}
for all bounded functions $f$ on $ \frac1n\NN_0^\X$ where $\Q\super{n}$ is the generator for $\rho\super{n}(t)$. Using the product structure of $\Pi\super{n}$ we have
\begin{equation}\label{eq:Pi perturbations}
    \Pi\super{n}(\rho+\tfrac1n\mathds1_x) = \Pi\super{n}(\rho) \tfrac{n\pi_x}{n\rho_x+1}, \ \ 
    \Pi\super{n}(\rho-\tfrac1n\mathds1_x) = \Pi\super{n}(\rho) \tfrac{\rho_x}{\pi_x}, \ \ 
    \Pi\super{n}(\rho+\tfrac1n\mathds1_x-\tfrac1n\mathds1_y) = \Pi\super{n}(\rho) (\tfrac{n\pi_x}{n\rho_x+1})(\tfrac{\rho_y}{\pi_y}). 
\end{equation}
Using this expression, and pulling out the function $f$,~\eqref{eq:abstract-stat} is equivalent to the following expression for any $\rho$
\begin{align*}
  &\sumxly \Bigl[ n(\rho_x+\tfrac1n) Q_{xy} \Pi\super{n}(\rho+\tfrac1n\mathds1_x-\tfrac1n\mathds1_y) - n\rho_x Q_{xy} \Pi\super{n}(\rho) \Bigr]\\
  &\qquad+\sum_x \Bigl[  n\lambda_{\In x} \Pi\super{n}(\rho-\tfrac1n\mathds1_x) - n\lambda_{\In x} \Pi\super{n}(\rho) \Bigr] 
  +\sum_{x\in\X} \Bigl[ n (\rho_x + \tfrac1n)\lambda_{\Out x} \Pi\super{n}(\rho+\tfrac1n\mathds1_x) - n\rho_x\lambda_{\Out x} \Pi\super{n}(\rho) \Bigr] \\
    &\stackrel{\eqref{eq:Pi perturbations}}{=} n\Pi\super{n}(\rho)\sum_{x\in\X} \tfrac{\rho_x}{\pi_x} \bigl[ 
    \underbrace{\sum_{\substack{ y\in\X\\ y\neq x}} (\pi_y Q_{yx} - \pi_x Q_{xy}) + \lambda_{\In x} - \pi_x\lambda_{\Out x} }_{=0}  \bigr]  + n\Pi\super{n}(\rho)\underbrace{\sum_x(\pi_x \lambda_{\Out x}-\lambda_{\In x}  )}_{=0},
\end{align*}
where both sums are $0$ since $\pi$ is the steady state of \eqref{eq:ODE}.

\paragraph{Properties of macroscopic steady state.}

If the graph is closed, i.e. $\lambda_{\In},\lambda_{\Out}=0$, then~\eqref{eq:ODE} is the Chapman-Kolmogorov equation for an irreducible Markov chain. Hence there is a coordinate-wise positive steady state, which is unique if the total mass $\sum_{x\in\X}\pi_x$ matches that of the initial condition $\rho(0)$~\cite[Thm.\ 3.5.2]{Norris1998}. 

We now show that there exists a unique coordinate-wise positive steady state irregardless of the initial condition even when the graph is not closed, but satisfies the assumptions made in Section~\ref{sec:model}.

Since the graph is not closed and irreducible there exists at least one $x$ such that $\lambda_{\In x},\lambda_{\Out x}>0$. This implies that the matrix $(Q-\diag(\lambda_{\Out}))$ is diagonally dominant with at least one strongly diagonally dominant row $\lvert Q_{xx}-\lambda_{\Out x}\rvert > \sum_{y\neq x} \lvert Q_{xy}\rvert$. Furthermore, the matrix is irreducible since the graph is assumed to be irreducible. These properties imply that $(Q-\diag(\lambda_{\Out}))$ is invertible \cite[Cor.~6.2.27]{HornJohnson1990} and so there exists a unique solution $\pi$ of 
\begin{equation}
    (Q-\diag(\lambda_{\Out}))\tp\pi=-\lambda_{\In}.
\label{eq:stability eq}
\end{equation}

To study the sign of $\pi$, we decompose the graph $\X$ into $\X^+\coloneqq\{\pi_x\geq0\}$ and $\X^-\coloneqq\{\pi_x<0\}$. If $\X^+=\emptyset$ then summing the stability equation~\eqref{eq:stability eq} over all of $\X=\X^-$ leads to the contradiction
\begin{equation*}
    0=\sum_{x\in\X^-}(\pi_x\lambda_{\Out x}-\lambda_{\In x})<0.
\end{equation*}
Similarly, if $\X^-,\X^+\neq\emptyset$, then summing the stability equation~\eqref{eq:stability eq} over $\X^-$ gives the contradiction
\begin{equation*}
    0=\sum_{x\in \X^-} \sum_{y\in \X^+} (\pi_x Q_{xy}-\pi_y Q_{yx}) + \sum_{x\in\X^-}(\pi_x\lambda_{\Out x}-\lambda_{\In x})<0
\end{equation*}
since by irreducibility there is at least one pair $x\in\X^-$, $y\in\X^+$ for which 
$Q_{xy}>0$, and all other terms are non-positive. We have thus shown that $\X=\X^+$.

Finally, to show that $\pi$ is coordinate-wise positive, i.e.\ $\pi_x>0$ for every $x$, assume by contradiction that there exists an $x\in\X$ for which $\pi_x=0$. Since that node does not have any outflow, the stability equation in $x$ reads
\begin{equation*}
    0=\sum_{y\neq x} (\pi_x Q_{xy}-\pi_y Q_{yx}) + \pi_x\lambda_{\Out x}-\lambda_{\In x}
    =-\sum_{y\neq x} \pi_y Q_{yx} -\lambda_{\In x},
\end{equation*}
and so $\lambda_{\In x}=0$ and $\pi_y=0$ whenever $Q_{yx}>0$. By irreducibility and recursion, this would lead to the contradiction $\lambda_{\In}=0$.

\subsection{Expressions for modified cost functions}
\label{app:modified cost functions}

Equations~\eqref{eq:Lsym-j0},~\eqref{eq:free energy loss} give expressions for the symmetric and antisymmetric cost evaluated at $j^0$. The general expressions for these costs are  
\begin{align*}
    \L^\sym(\rho,j)&=\sumxly \inf_{j^+_{xy}\geq0} \Bigl[ s\big(j^+_{xy} \mid \rho_x\sqrt{Q_{xy}Q_{yx}\mfrac{\pi_y}{\pi_x}}\big) + s\big(j^+_{xy}-j_{xy} \mid \rho_y\sqrt{Q_{xy}Q_{yx}\mfrac{\pi_x}{\pi_y}}\big)\Bigr] \notag \\
    &\qquad\quad + \sum_{x\in\X} \inf_{j^+_{\In x}\geq0} \Bigl[   s\big(j^+_{\In x} \mid \sqrt{\lambda_{\In x} \pi_x\lambda_{\Out x}}\big) + s\big(j^+_{\In x}-j_{\In x} \mid \rho_x\sqrt{\mfrac{\lambda_{\In x}\lambda_{\Out x}}{\pi_x}}\big)\Bigr],\\
    \L^\asym(\rho,j)&=\sumxly \inf_{j^+_{xy}\geq0} \Bigl[ s\big(j^+_{xy} \mid \sqrt{\rho_x\rho_y}Q_{xy}\sqrt{\mfrac{\pi_x}{\pi_y}}\big) + s\big(j^+_{xy}-j_{xy} \mid\sqrt{\rho_x\rho_y}Q_{yx}\sqrt{\mfrac{\pi_y}{\pi_x}}\big)\Bigr] \notag \\
    &\qquad\quad + \sum_{x\in\X} \inf_{j^+_{\In x}\geq0} \Bigl[   s\big(j^+_{\In x} \mid \lambda_{\In x}\sqrt{\mfrac{\rho_x}{\pi_x}}\big) + s\big(j^+_{\In x}-j_{\In x} \mid \sqrt{\rho_x\pi_x\lambda_{\Out x}}\big)\Bigr],
\end{align*}
with the corresponding Hamiltonians 
\begin{align*}
    \H^\sym(\rho,\zeta)&=\sumxly \Bigl[ \rho_x\sqrt{Q_{xy}Q_{yx}\tfrac{\pi_y}{\pi_x}}\bigl(e^{\zeta_{xy}}-1\bigr)+\rho_y\sqrt{Q_{xy}Q_{yx}\tfrac{\pi_x}{\pi_y}}\bigl(e^{-\zeta_{xy}}-1\bigr)\Bigr]\\
    &\qquad\quad + \sum_{x\in\X} \Bigl[  \sqrt{\lambda_{\In x}\pi_x\lambda_{\Out x}}\bigl(e^{\zeta_{\In x}}-1\bigr) + \rho_x\sqrt{\tfrac{\lambda_{\In x}\lambda_{\Out x}}{\pi_x}}\bigl(e^{-\zeta_{\In x}}-1\bigr)  \Bigr],\\
    \H^\asym(\rho,\zeta)&=\sumxly \Bigl[ \sqrt{\rho_x\rho_y}Q_{xy}\sqrt{\tfrac{\pi_x}{\pi_y}}\bigl(e^{\zeta_{xy}}-1\bigr)+\sqrt{\rho_x\rho_y}Q_{yx}\sqrt{\mfrac{\pi_y}{\pi_x}}\bigl(e^{-\zeta_{xy}}-1\bigr)\Bigr]\\
    &\qquad\quad + \sum_{x\in\X} \Bigl[  \lambda_{\In x}\sqrt{\tfrac{\rho_x}{\pi_x}}\bigl(e^{\zeta_{\In x}}-1\bigr) + \sqrt{\rho_x\pi_x\lambda_{\Out x}}\bigl(e^{-\zeta_{\In x}}-1\bigr)  \Bigr].   
\end{align*}
The integral $\int_0^T\!\L^\sym(\rho(t),j(t))\,dt$ is the large-deviation rate functional for the particle density and flux of a modified system, where particles jump from $x$ to $y$ with jump rate 
$n\rho_x\sqrt{Q_{xy}Q_{yx}\pi_y/\pi_x}$, particles are created at $x$ with rate $\sqrt{\lambda_{\In x} \pi_x\lambda_{\Out x}}$ and destroyed with rate $n\rho_x\sqrt{\lambda_{\In x}\lambda_{\Out x}/\pi_x}$. Similarly, $\int_0^T\!\L^\asym(\rho(t),j(t))\,dt$ corresponds to a system where particles jump from $x$ to $y$ with jump rate $n\sqrt{\rho_x\rho_y}Q_{xy}\sqrt{\pi_x/\pi_y}$, particles are created at $x$ with rate $n\lambda_{\In x}\sqrt{\rho_x/\pi_x}$ and destroyed with rate $n\sqrt{\rho_x\pi_x\lambda_{\Out x}}$. Observe that the symmetrised system describes independent jumping and destruction and constant creation as in the original system, whereas the antisymmetrised system introduces a nonlinear interaction between the particles.

\subsection{Derivation of Hamiltonian structure}
\label{app:Hamilton}

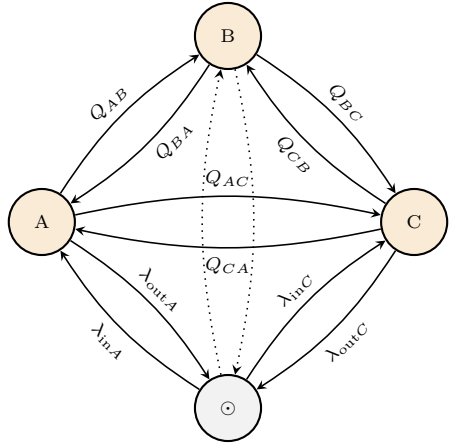
\begin{figure}[thb]
\centering
\begin{tikzpicture}[>=stealth]
\tikzstyle{every node}=[font=\scriptsize];

\node[state, fill=antiquewhite] (B) {B};
\node[state, below right of=B, fill=antiquewhite] (C) {C};
\node[state, below left of=B, fill=antiquewhite] (A) {A};
\node[state, below right of=A] (ghost) {$\ghost$};


\draw (B) edge[->,bend left=12] node[above,pos=0.5,sloped] {$Q_{BC}$}  (C);
\draw (A) edge[->,bend left=12] node[above,pos=0.5,sloped] {$Q_{AB}$} (B);
\draw (C) edge[->,bend left=12] node[below,pos=0.5,sloped] {$Q_{CB}$} (B);
\draw (A) edge[->,bend left=12] node[above,pos=0.5,sloped] {$Q_{AC}$} (C);
\draw (B) edge[->,bend left=12] node[below,pos=0.4,sloped] {$Q_{BA}$} (A);
\draw (C) edge[->,bend left=12] node[below,pos=0.5,sloped] {$Q_{CA}$} (A);

\draw (A) edge[->,bend left=12] node[above,pos=0.5,sloped] {$\lambda_{\Out A}$} (ghost);
\draw (ghost) edge[->,bend left=12] node[below,pos=0.5,sloped] {$\lambda_{\In A}$} (A);
\draw (C) edge[->,bend left=12] node[below,pos=0.5,sloped] {$\lambda_{\Out C}$} (ghost);
\draw (ghost) edge[->,bend left=12] node[above,pos=0.5,sloped] {$\lambda_{\In C}$} (C);
\draw (B) edge[->,dotted,bend left=12] (ghost);
\draw (ghost) edge[->,dotted,bend left=12] (B);

\end{tikzpicture}
\caption{The graph from Figure~\ref{fig:open-bound-chain} with an additional ghost node}
\label{fig:ghost node}
\end{figure}

We expand the graph with an additional ghost node $\tilde\X:=\X\cup\{\ghost\}$, where mass flowing in and out of the system is now extracted from respectively collected in $\ghost$ instead, see Figure~\ref{fig:ghost node}. This results in a dynamics that conserves the total mass $M:=\sum_{x\in\X\cup{\ghost}}\rho_x(0)$ (although $\rho_{\ghost}(t)$ may become negative), and the rates of flowing out of a node $x$ is either linear $\rho_xQ_{xy}$, $\rho_x\lambda_{\Out x}$ or constant $\lambda_{\In x}$. The expanded system has the same, coordinate-wise positive steady state $\pi$ on $\X$ as the original system, but with an additional coordinate $\pi_\ghost$. By mass conservation, this coordinate satisfies $\pi_\ghost=M-\sum_{x\in\X}\pi_x$, so if we initially place enough mass in the ghost node (which does not change the dynamics), then $M$ will be sufficiently large so that $\pi_\ghost>0$. 

We are then in the same setting as zero-range processes~\cite[Prop.~5.3]{PRS2021TR}. By results therein, the augmented antisymmetric zero-cost dynamics is a Hamiltonian flow and can be written as (abbreviating $\rho_\X:=(\rho_x)_{x\in\X}$ defined in~\eqref{eq:asym ODE})
\begin{equation}
    \begin{bmatrix}
      \dot\rho_\X(t)\\
      \dot\rho_{\ghost}(t)
    \end{bmatrix}
    =
    \tilde\JJ(\rho)\grad\tilde\U(\rho)
    \coloneqq
    \begin{bmatrix}
        \JJ(\rho)      & \JJ_{\X\ghost}(\rho)\\
        -\JJ_{\X\ghost}(\rho) & \JJ_{\ghost\ghost}(\rho)
    \end{bmatrix}
    \begin{bmatrix}
        \grad_{\rho_\X}\tilde\U(\rho)\\
        \grad_{\rho_{\ghost}}\tilde\U(\rho)
    \end{bmatrix},
\label{eq:expanded Ham structure}
\end{equation}
where $\tilde\U(\rho,\rho_\ghost)=\U(\rho)$ and $\U,\JJ$ are given by \eqref{eq:Hamiltonian energy},\eqref{eq:Poisson structure} and $\JJ_{\X\ghost},\JJ_{\ghost\ghost}$ are irrelevant by the following argument. Mass conservation implies that $\Lambda:(\rho_\X,\rho_\ghost)\mapsto\rho_\X$ is a bijection with Jacobian $J_\Lambda=\lbrack I \mid 0\rbrack$. Applying the variable transformation $\Lambda$ to \eqref{eq:expanded Ham structure} yields $\dot\rho_\X(t)=J_\Lambda\tilde\JJ(\rho) J_\Lambda\tp \grad_{\rho_\X}\U(\rho)=\JJ(\rho)\grad_{\rho_\X}\U(\rho)$ as claimed.

\subsection{Jacobi identity}
\label{app:Jacobi}

We verify that the bracket $\lbrack \mathscr F^1,\mathscr F^2\rbrack_\rho=\grad\mathscr F^1(\rho)\cdot\JJ(\rho)\grad\mathscr F^2(\rho)$ defined by the Poisson structure~\eqref{eq:Poisson structure} indeed satisfies the Jacobi identity 
$
\lbrack\lbrack \mathscr F^1,\mathscr F^2\rbrack,\mathscr F^3\rbrack_\rho
+
\lbrack\lbrack \mathscr F^2,\mathscr F^3\rbrack,\mathscr F^1\rbrack_\rho
+
\lbrack\lbrack \mathscr F^3,\mathscr F^1\rbrack,\mathscr F^2\rbrack_\rho=0$ for all sufficiently smooth functions $\mathscr F^i$ and all $\rho\in\RR^\X$. Omitting $\rho$-dependencies to shorten notation, this identity is equivalent to the following tensor relation~\cite[Lem~A.1]{PRS2021TR}, for all $\rho\in\RR^\X$ and $x,y,z\in\X$,
\begin{align}
    R^y_{xz}+R^z_{yx}+R^x_{zy}\equiv0,
&&
    R_{xy}^z\coloneqq\sum_{a\neq z}\JJ_{az}\partial_a\JJ_{xy}.
\label{eq:alt Jacobi}
\end{align}
We first calculate the derivative for $x\neq y$ (clearly $\JJ_{xx}\equiv0$),
\begin{align*}
    \partial_a\JJ_{xy}&=
        \begin{cases}
            \sqrt{\mfrac{\rho_x\rho_y}{\rho_a}} B_{xy}^a, &a\neq x,y\\
            \sum_{z\neq x} \sqrt{\mfrac{\rho_y\rho_z}{\rho_x}} B_{xy}^z + 2\sqrt{\rho_y} B_{xy}^x+\sqrt{\mfrac{\rho_y}{\rho_x}}B_{xy}^\ghost, &a=x,\\
            \sum_{z\neq y} \sqrt{\mfrac{\rho_x\rho_z}{\rho_y}} B_{xy}^z + 2\sqrt{\rho_x} B_{xy}^y+\sqrt{\mfrac{\rho_x}{\rho_y}}B_{xy}^\ghost, &a=y,
        \end{cases}\\
    B_{xy}^z&\coloneqq \sqrt{\mfrac{\pi_x\pi_z}{\pi_y}}Q_{zy}-\sqrt{\mfrac{\pi_x\pi_y}{\pi_z}}Q_{yz}-\sqrt{\mfrac{\pi_y\pi_z}{\pi_x}}Q_{zx}+\sqrt{\mfrac{\pi_x\pi_y}{\pi_z}}Q_{xz},\\
    B_{xy}^\ghost&\coloneqq \sqrt{\mfrac{\pi_x}{\pi_y}}\lambda_{\In y}-\sqrt{\pi_x\pi_y}\lambda_{\Out y}-\sqrt{\mfrac{\pi_y}{\pi_x}}\lambda_{\In x} + \sqrt{\pi_x\pi_y}\lambda_{\Out x}.
\end{align*}
The tensor then decomposes into terms of different orders $R_{xy}^z=\tensor*[^2]{R}{^z_{xy}}+\tensor*[^3]{R}{^z_{xy}}+\tensor*[^4]{R}{^z_{xy}}$ of $\sqrt{\rho}$, where
\begin{align*}
  \tensor*[^2]{R}{^z_{xy}}&\coloneqq 2\big\lbrack \sqrt{\rho_x\rho_z} B_{xy}^\ghost B_{yz}^\ghost
    +\sqrt{\rho_y\rho_z} B_{xy}^\ghost B_{xz}^\ghost\big\rbrack,\\
  \tensor*[^3]{R}{^z_{xy}}&\coloneqq 2\sum_a \big\lbrack \sqrt{\rho_x\rho_y\rho_z} B_{xy}^a B_{az}^\ghost
    + \sqrt{\rho_a\rho_y\rho_z} (B_{xy}^a B_{xz}^\ghost + B_{xy}^\ghost B_{xz}^a)
    + \sqrt{\rho_a\rho_x\rho_z} (B_{xy}^a B_{yz}^\ghost + B_{xy}^\ghost B_{yz}^a)\big\rbrack,\\
  \tensor*[^4]{R}{^z_{xy}}&\coloneqq 2\sumsum_{a,b} \big\lbrack \sqrt{\rho_b\rho_x\rho_y\rho_z} B_{xy}^a B_{az}^b
    + \sqrt{\rho_a\rho_b\rho_y\rho_z} B_{xy}^a B_{xz}^b
    + \sqrt{\rho_a\rho_b\rho_x\rho_z} B_{xy}^a B_{yz}^b \big\rbrack.
\end{align*}
Since \eqref{eq:alt Jacobi} needs to hold for all $\rho\in\RR^\X$, we may check it for each order separately. Using the skew-symmetry of $(B_{xy}^\ghost)_{xy}$, for the second-order terms we have
\begin{align*}
  \tensor*[^2]{R}{^y_{xz}}+\tensor*[^2]{R}{^z_{yx}}+\tensor*[^2]{R}{^x_{zy}}=
    2\sqrt{\rho_x\rho_y} B_{zy}^\ghost \big\lbrack B_{xz}^\ghost+B_{zx}^\ghost\big\rbrack
    +2\sqrt{\rho_x\rho_z} B_{yx}^\ghost \big\lbrack B_{yz}^\ghost+B_{zy}^\ghost\big\rbrack
    +2\sqrt{\rho_y\rho_z} B_{xz}^\ghost \big\lbrack B_{xy}^\ghost+B_{yx}^\ghost\big\rbrack\equiv0
\end{align*}
%
Using the skew-symmetry of $(B_{xy}^z)_{xy}$ and $(B_{xy}^\ghost)_{xy}$, for the third order terms we find
\begin{align*}
  \tensor*[^3]{R}{^y_{xz}}+\tensor*[^3]{R}{^z_{yx}}+\tensor*[^3]{R}{^x_{zy}}&=
2\textstyle\sum_a\big\lbrack
   \sqrt{\rho_x\rho_y\rho_z}\big( B_{xz}^a B_{ay}^\ghost + B_{yx}^a B_{az}^\ghost+B_{zy}^a B_{ax}^\ghost\big)\\
  &\qquad\quad+\sqrt{\rho_a\rho_x\rho_y}\big( B_{xz}^a B_{zy}^\ghost + B_{zx}^a B_{zy}^\ghost + B_{zy}^a B_{xz}^\ghost+B_{zy}^a B_{zx}^\ghost\big)\\
  &\qquad\quad+\sqrt{\rho_a\rho_x\rho_z}\big( B_{yx}^a B_{yz}^\ghost + B_{yx}^a B_{zy}^\ghost + B_{yz}^a B_{yx}^\ghost+B_{zy}^a B_{yx}^\ghost\big)\\
  &\qquad\quad+\sqrt{\rho_a\rho_y\rho_z}\big( B_{xz}^a B_{xy}^\ghost + B_{xz}^a B_{yx}^\ghost + B_{xy}^a B_{xz}^\ghost+B_{yx}^a B_{xz}^\ghost\big) \big\rbrack\\
  &=2\sqrt{\rho_x\rho_y\rho_z}\sum_a \big( B_{xz}^a B_{ay}^\ghost + B_{yx}^a B_{az}^\ghost+B_{zy}^a B_{ax}^\ghost\big).
\end{align*}
Hence the sum over the constants needs to be zero. After a lengthy calculation we find
\begin{align*}
    &\sum_a \big( B_{xz}^a B_{ay}^\ghost + B_{yx}^a B_{az}^\ghost+B_{zy}^a B_{ax}^\ghost\big)\\
    &\quad=\mfrac1{\sqrt{\pi_z}}\Big( \sqrt{\mfrac{\pi_x}{\pi_y}}\lambda_{\In y} - \sqrt{\pi_x\pi_y}\lambda_{\Out y} - \sqrt{\mfrac{\pi_y}{\pi_x}}\lambda_{\In x} + \sqrt{\pi_x\pi_y}\lambda_{\Out x}\Big) \sum_{a\neq z}\big(\pi_a Q_{az}-\pi_z Q_{za}\big)\\
    &\qquad+\mfrac1{\sqrt{\pi_x}}\Big( \sqrt{\mfrac{\pi_y}{\pi_z}}\lambda_{\In z} - \sqrt{\pi_y\pi_z}\lambda_{\Out z} - \sqrt{\mfrac{\pi_z}{\pi_y}}\lambda_{\In y} + \sqrt{\pi_y\pi_z}\lambda_{\Out y}\Big) \sum_{a\neq z}\big(\pi_a Q_{ax}-\pi_x Q_{xa}\big)\\
    &\qquad+\mfrac1{\sqrt{\pi_y}}\Big( \sqrt{\mfrac{\pi_z}{\pi_x}}\lambda_{\In x} - \sqrt{\pi_x\pi_z}\lambda_{\Out x} - \sqrt{\mfrac{\pi_x}{\pi_z}}\lambda_{\In z} + \sqrt{\pi_x\pi_z}\lambda_{\Out z}\Big) \sum_{a\neq z}\big(\pi_a Q_{ay}-\pi_y Q_{ya}\big). 
\end{align*}
Using the stability equation \eqref{eq:stability eq}, the three sums on the right can be replaced by expressions depending on $\lambda_\In,\lambda_\Out$ only. This yields twelve paired terms that cancel each other out, so that indeed $\tensor*[^3]{R}{^y_{xz}}+\tensor*[^3]{R}{^z_{yx}}+\tensor*[^3]{R}{^x_{zy}}\equiv0$.

Finally, for the fourth order terms, $\tensor*[^4]{R}{^y_{xz}}+\tensor*[^4]{R}{^z_{yx}}+\tensor*[^4]{R}{^x_{zy}}\equiv0$, because this describes the closed graph setting $\lambda_\In,\lambda_\Out=0$, which satisfies the Jacobi identity~\cite[App.~A]{PRS2021TR}.

{\small
\bibliography{bib}
\bibliographystyle{alphainitials}
}


%

\end{document}